\documentclass{article}
\usepackage{amsmath}
\usepackage{hyperref}
\usepackage{moreverb}
\usepackage[normalem]{ulem}
\usepackage{graphicx}
\usepackage{tabularx}
\usepackage{hypcap}
\usepackage{authblk}
\usepackage{float}
\usepackage[a4paper, left=2.5cm, right=2.5cm, top=2.5cm]{geometry}
\usepackage[backend=biber]{biblatex}
\graphicspath{{./figures/}}

\newcommand\BibTeX{{\rmfamily B\kern-.05em \textsc{i\kern-.025em b}\kern-.08em
T\kern-.1667em\lower.7ex\hbox{E}\kern-.125emX}}

\begin{document}

\title{Design Considerations for a Phase II platform trial in Major Depressive Disorder}

\author[1]{Michaela Maria Freitag} 
\author[1]{Dario Zocholl}
\author[2,3]{Elias Laurin Meyer} 
\author[4,5,6,7]{Stefan M. Gold} 
\author[2]{Marta Bofill Roig}
\author[8]{Heidi De Smedt}
\author[2]{Martin Posch}
\author[2]{Franz König \thanks{Correspondence to franz.koenig@meduniwien.ac.at}}
\author[9]{on behalf of the EU-PEARL MDD Investigators}

\affil[1] {\small Charité – Universitätsmedizin Berlin, corporate member of Freie Universität Berlin, Humboldt-Universität zu Berlin, and Berlin Institute of Health, Institute of Biometry and Clinical Epidemiology, Berlin, Germany}
\affil[2]{\small Medical University of Vienna, Center for Medical Data Science, Vienna, Austria}
\affil[3]{\small Berry Consultants, Vienna, Austria}
\affil[4]{Charité – Universitätsmedizin Berlin, corporate member of Freie Universität Berlin, Humboldt-Universität zu Berlin, and Berlin Institute of Health, Department of Psychiatry and Psychotherapy, Berlin, Germany}
\affil[5]{Charité – Universitätsmedizin Berlin, corporate member of Freie Universität Berlin, Humboldt-Universität zu Berlin, and Berlin Institute of Health, Medical Department, Section Psychosomatics, Berlin, Germany}
\affil[6]{DZPG, German Center for Mental Health}
\affil[7]{Universitätsklinikum Hamburg-Eppendorf, INIMS, Hamburg, Germany}
\affil[8]{Janssen Research and Development, Beerse, Belgium}
\affil[9]{Membership of The EU-PEARL MDD Investigators is provided in the Acknowledgments}


\maketitle

\section{Abstract}
Major Depressive Disorder (MDD) is one of the most common causes of disability worldwide. Unfortunately, about one-third of patients do not benefit sufficiently from available treatments and not many new drugs have been developed in this area in recent years. We thus need better and faster ways to evaluate many different treatment options quickly. Platform trials are a possible remedy - they facilitate the evaluation of more investigational treatments in a shorter period of time by sharing controls, as well as reducing clinical trial activation and recruitment times.
We discuss design considerations for a platform trial in MDD, taking into account the unique disease characteristics, and present the results of extensive simulations to investigate the operating characteristics under various realistic scenarios. To allow the testing of more treatments, interim futility analyses should be performed to eliminate treatments that have either no or negligible treatment effect. 
Furthermore, we investigate different randomisation and allocation strategies as well as the impact of the per-treatment arm sample size. We compare the operating characteristics of such platform trials to those of traditional randomised controlled trials and highlight the potential advantages of platform trials.


\section{Introduction}\label{sec1}
Major depressive disorder (MDD) is one of the leading causes of disability. Studies based on clinical interviews indicate an estimated worldwide prevalence of 5--6 \% at any given time and a 1 in 6 lifetime prevalence \cite{Otte}. Additionally, MDD is associated with a twofold increased risk to develop other medical diseases and about 8-10 life years lost compared to the general population \cite{lancet}. While there are several different antidepressant treatments available, about 50\% of patients do not benefit sufficiently from the first treatment ("Partially Responsive Depression", PRD) and the majority of these also do not benefit from second-line treatment ("Treatment-Resistant Depression", TRD) \cite{trd}. Up to one-third of MDD patients do not achieve full symptomatic remission despite multiple medication attempts \cite{rush}. Moreover, in recent years \cite{fda_approval}, comparatively few new drugs have been developed and approved for psychiatric indications, including MDD, especially few with new mechanisms of action. Faster and more efficient development procedures, including novel trial designs, could thus contribute to facilitating drug development, especially for early-phase studies in this field. 
Besides multi-arm multi-stage trials and group sequential designs, one option to improve the efficiency of study designs is the use of master protocols and platform trials in particular. Such designs can a) accelerate the development and approval of new treatments by enabling the investigation of multiple treatments in parallel, sharing controls and reducing clinical trial activation times as well as recruitment times and b) lead to lower costs and higher quality data while being more patient-centric \cite{collignon2021,meyer2020}. Some successful adaptive platform trials have already been implemented in the past, e.g. REMAP-CAP in Covid and lung diseases \cite{remap} and I-SPY 2 in breast cancer \cite{ispy, adaptiveplatform}. 
There are multiple definitions of platform trials \cite{hirakawa, renfro2017, meyer2020}. This article considers them as clinical trials allowing for simultaneous and sequential evaluation of multiple interventions in one indication against a common control possibly taking into account specific disease sub-types. Their unique feature is the possibility of treatments joining or leaving the trial over time \cite{woodcock, meyer2020}. This definition is consistent with the one used in the FDA guidance document on master protocols \cite{fda}.
The IMI project EU-PEARL (EU Patient-cEntric clinicAl tRial pLatforms) aims to promote the development of platform trials by providing a generic framework to design and conduct integrated research platforms\cite{koenig2023, eu-pearl}. The project is motivated by use cases in four indications still facing high unmet needs, one of them being MDD. 
The proposed platform trial design in MDD was created in an iterative process involving different stakeholders such as clinicians, statisticians, pharmacists, and experts by experience. Additionally, a meeting with EMA representatives took place to discuss several design elements under consideration. The goal was to develop an efficient phase II trial design to screen novel treatments and repurposed drugs for use in MDD. The two subpopulations of MDD to be investigated are patients with PRD or TRD. Potential treatments may work in both groups but may also differ with regard to their efficacy and safety between these groups. With the caveat that TRD is a more severe subpopulation than PRD, and thus a treatment that is effective for PRD may not have the same effectiveness in TRD. For the purpose of this simulation, we have made the assumption that both would show identical results.
Therefore we focus on presenting EU-PEARL considerations for designing and simulating platform trials in TRD. 
The summaries of the platform design considerations for one of the other use cases, Non-Alcoholic Steatohepatitis (NASH), have already been published \cite{meyer2022decision, meyer2023designing} alongside a general master protocol template \cite{protocol-template}. 
\\ 
For a platform trial, a number of trial characteristics need to be tailored to the purpose of the trial, including choice of endpoint, sample size, definition of control group (concurrent vs non-concurrent), allocation ratio, etc. In the current paper, we explore the impact of these design choices on trial performance and derive informed recommendations for a phase II platform trial in TRD.
\\
Platform trials increase efficiency due to e.g. shared controls and infrastructure. The aim is to maintain some of the flexibility that is available in separate trials also within the platform trial, but it will be limited as e.g. endpoints should be the same in all arms. The gained efficiency comes with statistical challenges and design challenges when planning platform trials. Adaptations, design, and analysis elements have to be carefully tailored specifically to the area of application.
For example, the sample size reduction realized by platform trials vs. running multiple separate trials in parallel is mainly due to sharing of common control group data. One important question is if only the concurrent control data should be used for the analysis of an arm, i.e. only the data of control patients who could have been randomised to the treatment arm in question, or if all control data should be used, i.e. also non-concurrent controls. If all control data was used, the power would be higher, but time trends could lead to distortions especially if platform trials run for many years. Several methods have recently been developed that include time as a covariate in the analysis to address the time trend effects \cite{bofill_trends, time_machine}. \\
Another statistical challenge is the implementation of adaptive interim analyses during which possible adaptations can be made. One such adaptation is the possibility of terminating treatment arms early and enabling faster decisions. In a platform trial, it is critical to drop non-promising treatments due to futility and continue with the promising ones. By dropping one arm, resources become available to test another arm. This is especially valuable in phase II settings when screening for potentially active treatments is emphasized. 
\\
In this paper, we develop a phase II platform trial design in TRD and investigate a broad range of design options including allocation strategies to different arms and dropping treatment arms for futility. In section 2 we define the hypotheses to be tested, the analysis methods, trial design options, and the setup of the simulation study. In section 3 we report the results of the simulation study for different design elements. We also compare the performance of the proposed platform trial with traditional 2-arm trials, illustrating the potential benefits of the platform trial approach. We close with a discussion in section 4.

\section{Methods}
\subsection{General Design Aspects}
\begin{figure*}[h]
\centering
  \includegraphics[width=11cm]{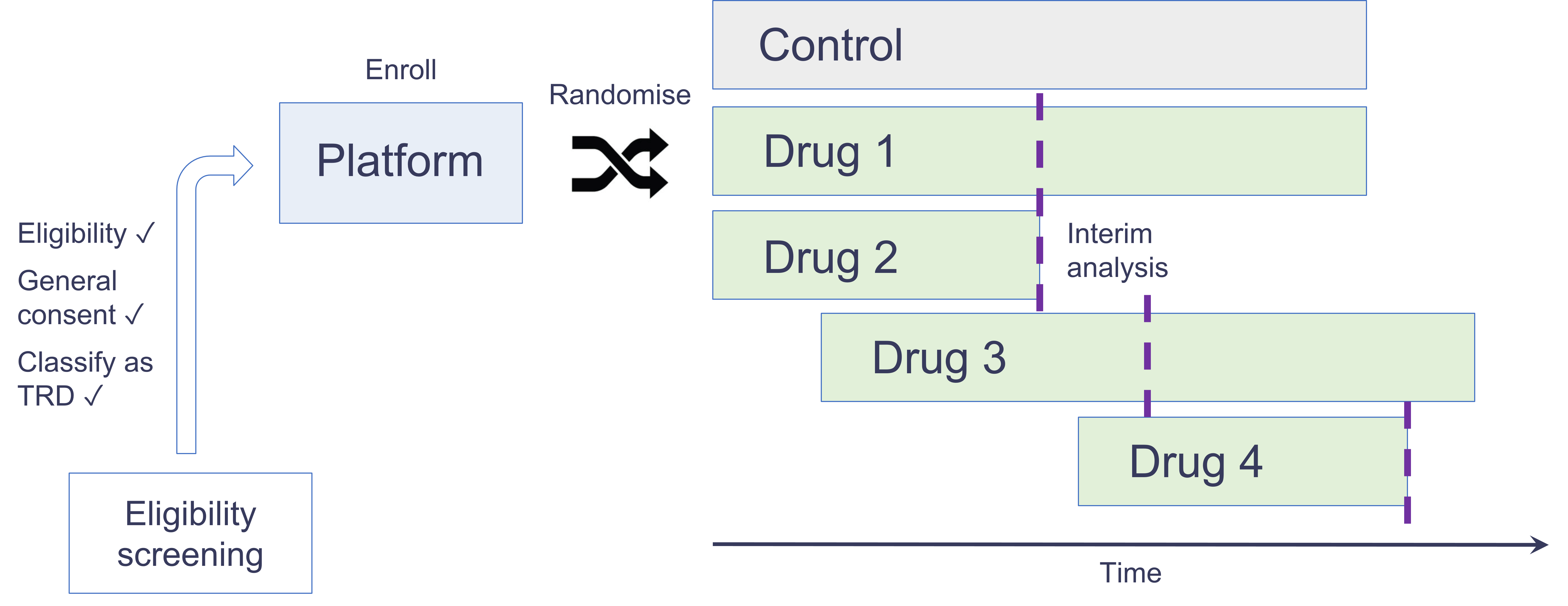}
    \caption{MDD platform trial design proposal. If patients are eligible for the platform, they are randomised to either control or one of the treatment arms within the platform. Different drugs can enter the trial at different time points. The design might also allow for interim analyses to drop treatment arms early. The time point of the interim analyses is indicated by a dotted line.} 
    \label{fig:MDD_design}
\end{figure*}

The target population for the platform trial are patients with Treatment-resistant Depression (TRD). If patients are eligible for the trial, they are invited to enrol and subsequently be randomised to either the control arm or a treatment arm. The allocated treatment (or placebo) is administered in addition to antidepressants and therapies the patients were receiving at the time of inclusion in the trial. As the primary outcome measure, we evaluate the change in Montgomery-Åsberg Depression Rating Scale (MADRS) score between week 6 and the baseline value. Both the duration of 6 weeks for short-term trials and the use of MADRS score are common standard elements in depression trials \cite{ema_depression}.  
For all analyses in the platform trial, we use a one-sided significance level of 0.05 and only concurrent controls, i.e. only information from patients in the control arm that could have been randomised to the treatment arm in question.
Figure \ref{fig:MDD_design} shows a schematic of the specific platform trial design for TRD patients. 
Furthermore, the allocation ratio to control is dependent on the actual number of enrolling treatment arms. We suggest a modified block randomisation. It is discussed in detail in section \ref{sec:block}.
For the specific design elements of the platform trial, we mainly focus on the selection of adequate allocation ratios, the selection of futility boundaries and the handling of time trends.
We will also consider standard 2-arm trials and compare their performance later on in the results section to the performance of platform trials. For good comparability, we selected the same general design aspects for both types of trials

\subsection{Statistical Methods}

\subsubsection{Hypotheses}
In a platform trial, several treatment arms may enter and leave the trial at different time points. The goal of the design under consideration is to compare the efficacy of each experimental treatment against a shared control. The arms are therefore tested individually without comparing the effect of one experimental treatment to that of another. The platform trial also does not investigate a global hypothesis. \\
Consider a platform trial where investigational treatments $j$, $j \in \{1,..., J\}$ are compared to a common control $c$. The objective of this platform trial is to find any efficacious treatment, i.e. any treatment that lowers the MADRS score at week 6 compared to a control treatment. The corresponding baseline value of the MADRS score will be measured at the time point patients are being randomized into either treatment arm $j$ or control arm $c$.
To demonstrate the efficacy of an experimental treatment $j$ against the control group $c$, an elementary null hypothesis $H^0_j$ is tested at a one-sided significance level $\alpha$. For any such test, control data that is concurrently collected will be used, thereby defining a different set $c_j$ of control data for every treatment $j$.
Let $\mu_j^{6w}$ and $\mu_j^{baseline}$ be the true means of the MADRS score of treatment $j$ at week 6 and baseline, respectively and $\mu_{c_j}^{6w}$ and $\mu_{c_j}^{baseline}$ the same values for the concurrent control group.
Then the one-sided null hypotheses are given by $H^0_j$: $\mu_{c_j}^{6w} - \mu_j^{6w} \leq 0$ vs the alternative $H_j^A$: $\mu_{c_j}^{6w} - \mu_j^{6w} > 0$ controlling for the baseline values. Alternatively, one could also express the null and alternative hypotheses for the change in MADRS score between week 6 and baseline with $\Delta_j=\mu_j^{baseline}-\mu_j^{6w}$ and $\Delta_{c_j}=\mu_{c_j}^{baseline}-\mu_{c_j}^{6w}$ for the experimental treatment $j$ and the concurrent control arm, respectively. Then the elementary null and alternative hypotheses being tested can be given as  
$$H^0_j: \Delta_{j} - \Delta_{c_j} \leq 0 \text{ vs } H_j^A: \Delta_{j} - \Delta_{c_j} > 0$$ 
\\

\subsubsection{Analysis Method}
For the analysis of the primary endpoint, i.e. the MADRS score at week 6 controlled for the baseline value, an analysis of covariance (ANCOVA) is conducted using the factor treatment and adjusting for the baseline MADRS score as a covariate. ANCOVA assumes linearity of the covariate effect and absence of covariate-by-group interaction. Adjusting for the baseline value has the advantage that the estimate and test of the treatment effect are equivalent when using the value at week 6 or the change between baseline and week 6 as the dependent variable in the ANCOVA model \cite{senn2021statistical, ema_baseline}. 
The idea of the ANCOVA is to use regression to control (i.e. adjust) for additional covariates such as the baseline value so one can study the post-treatment measure free of the proportion of variance linearly associated with the baseline.
The ANCOVA is also the preferred analysis method for randomised clinical trials with a pre- and post-treatment measurement in case of continuous endpoints and homogeneous covariance matrices \cite{winkens, zhang, breukelen}. 
\\
As mentioned before we test the different null hypotheses $H^0_j$ separately for every treatment $j$, $j \in \{1,..., J\}$. We therefore fit different models for every treatment arm $j$ using the data of treatment arm $j$ and the corresponding concurrent control data $c_j$ only. 
The formal models $M^j$, $j \in \{1, ..., J\}$ can be written as 
\begin{equation}
   M^j: Y_{ih_j} = \alpha_j + \beta_{j} G_{ih_j}  + \gamma_j X_{ih_j}+ e_{ih_j}, 
\end{equation}
whereby $Y_{ih_j}$ is the post-treatment MADRS score of person $i$ in group $h_j$ and $X_{ih_j}$ the corresponding baseline value. In every model, the index $h_j$ only takes two different values, $j$ for the treatment arm in question and $c_j$ for the concurrent control group.
For each ANOCOVA model $M^j$ the corresponding intercept is denoted by $\alpha_j$.
$G_{ih_j}$ is the treatment indicator for person $i$ (i.e. $G_{ih_j}=0$ for control and $G_{ih_j}=1$ for treatment), $\beta_{j}$ quantifies the treatment effect and $\gamma_{j}$ the baseline effect. 
Additionally a random term $e_{ih_j}$ is added, which is normally distributed with zero mean and constant variance. 
Practical use of ANCOVA requires estimation of $\gamma_{j}$, which is a function of the within-group variances and correlation of the pre-treatment and post-treatment scores. 
For the patients $i$ only the data of the $n_j$ subjects randomised to the experimental treatment arm $j$ of interest and the corresponding concurrent control data of $n_{c_j}$ subjects in the control group is used.
Please note that the size $n_{c_j}$ of concurrent control data might distinctly differ between experimental treatments $j$.
\\
Additionally, the ANCOVA model can be expanded to adjust for further variables. In the context of a platform trial, it is also reasonable to adjust not only for the MADRS score at baseline but also for the factor time period to avoid bias in the estimates due to time trends and to address the change in allocation ratios due to entering or dropping of arms. 
We define the categorical covariate time period as the time intervals where there are no arms entering or leaving the trial. So the time period changes whenever a treatment arm enters or leaves the platform trial. We explored a second ANCOVA model with the factor time period. This modelling allows control of the type 1 error under certain assumptions if there were changes in the treatment effects over time. We refer to Bofill et al.  \cite{bofill_trends} for a formal definition of time periods in the context of platform trials and for more detailed methods on how to adjust for potential time trends.


\subsubsection{Effect size definition} \label{sec:effects}
The effect sizes used in this paper are adjusted for the correlation between the baseline value and the value at week 6. They follow a logic similar to the calculation of Cohen's d, see formula \ref{eq:d}. $\Delta_C$ is the difference between baseline and week 6 in the control group and $\Delta_T$ is the corresponding value in the treatment group. 
\begin{eqnarray}
d=\frac{\Delta_T - \Delta_C}{\sqrt{\frac{\left(SD_{\Delta_T}^2 + SD_{\Delta_C}^2\right)}{2}}}.
\label{eq:d}
\end{eqnarray}
For the simulation we assume $SD_{\Delta_C}$ and $SD_{\Delta_T}$ to be equal. 
The value used in the simulation for the standard deviations is based on a variance-covariance matrix from data of a placebo control arm in a past phase II study in augmentation treatments in MDD.
Based on the same data, it is assumed that the MADRS score reduces  from 32 points at baseline to 20 at week 6 in the control group and that the correlation between baseline value and week 6 is 0.214. 
According to equation \ref{eq:d} a value of $d=0.2$ corresponds to an absolute reduction of 2.25 in the MADRS score compared to the control group, $d=0.35$ corresponds to an absolute reduction of 4, and $d=0.5$ of 5.7. \\
We assume a standardized effect size $d=0.35$ to be the 
clinically relevant effect for augmentation strategies in TRD. 
An effect of 0.5 is regarded as rather big. 

\subsubsection{Allocation ratios and randomisation methods} \label{sec:block}
There are many different options to define allocation rates to different treatment arms and the control arm in multi-arm trials like the platform trial. Common ones are a 1:1:...:1 allocation where every treatment arm and the control arm are allocated the same proportion of patients and a 1:1:..:1:$x$ allocation where $x$ relates to the ratio in the control arm and all treatment arms get the same number of patients but the control arm is allocated a different fraction.
The value $x$ can either be constant or dependent on the number $k$ of treatment arms concurrently enrolling in the trial. 
For multi-arm trials a square-root allocation, i.e. 1:1:...:1:$\sqrt{k}$, yields good results \cite{dunnett_sqrt}. It minimizes the standard error of treatment effect estimates for normally distributed endpoints with equal variances across groups. However, the adding and terminating of treatment arms during the course of a platform trial impacts the performance of allocation rules and other rules than the one for normal multi-arm trials may offer the best results \cite{bofill_allocation}. The allocation rate to control is not fixed upfront, but varies over time, depending on the number of treatments concurrently under evaluation, and the timing of entry and departure of interventions. 
It is to be noted that the (placebo) response in patients with MDD strongly depends on factors like the expectancy to receive placebo, i.e. it depends on the number of treatment arms that are recruiting at the same time \cite{papakostas}. In order to avoid large variations in the treatment effect over time, the allocation rate to control needs to be controlled. Therefore, it can be reasonable to consider a cap for minimal allocation to the control arm. Throughout the project, this cap was discussed multiple times with clinical experts and finally it was recommended that about one-third of patients should be randomised to the control arm. For the simulations presented in the results section, we used a cap of 35\%. Other values for the cap were also investigated and are presented in the online supplement.
\\
All considered allocation ratios can be easily achieved by using simple randomisation for any $x$ by modifying the randomisation probabilities accordingly. In order to limit the variability introduced by simple randomisation, we also implemented a modified version of block randomisation. A performance comparison of simple and block randomisation is provided in the online supplement.
\\
For the modified block randomisation we were aiming to get blocks of minimal size to reach the targeted allocation ratio of 1:1:...:1:$x$, where x corresponds to the ratio of the control arm.
For the 1:1:...:1 allocation, basic permuted blocks are created with the number of open treatment arms (including control) as the length of the block to be permuted, i.e. number of currently enrolling treatment arms (=$k$) + 1. For example, aiming at an allocation ratio of 1:1:1:1 a block with the spots $T_1 T_2 T_3 C$ is permuted.  
\\
For the other allocation methods aiming at a 1:1:..:1:$x$ ratio, additional control spots have to be added to the block with length $k+1$ to reach the desired ratio. Here $x$ denotes the control ratio. The length of the control blocks is equivalent to the number of patients still needed in the control group. 
If $x$ is an integer, then the final block length is simply $k+x$, with $x$ spots for the control in a permuted block. For example for the allocation ratio of 1:1:1:3 a block with the spots $T_1 T_2 T_3 C C C$ is permuted.  
If $x$ is not an integer, we combine block randomisation with some random elements on how many spots should eventually be added. So the basic block has $k$ spots for experimental treatments and $y$ spots for the control arm, with $y=\lfloor x \rfloor$. Furthermore, for each spot, it is decided whether to add an additional spot for the control arm with a probability of $Frac(x)$ or not.
For example for 1:1:1:$\sqrt{3}$ allocation with $x=\sqrt{3}=1.73$, this method results in a minimal block length of 4.
With a probability of $1-0.73=0.27$ such a permuted block $T_1 T_2 T_3 C$ with length 4 is taken or with probability $0.73$ a block $T_1 T_2 T_3 C C$ with length 5 is taken adding an additional spot for control. By applying this principle to randomly add an additional spot for control the targeted allocation ratio is approximately reached implementing a minimal block length in the randomisation procedure. This modified procedure combines elements from traditional block randomisation and random elements like in simple randomisation. 


\subsection{Simulations}\label{sec:sim}
During the design stage and simulation of any randomised controlled trial, several design specifications and assumptions are needed to understand the behaviour and operating characteristics of the trial. Design choices are aspects that can be controlled by the decision maker and assumptions refer to unknown quantities that have to be estimated from preexisting data \cite{benda, friede}. For a trial with continuous pre-treatment and post-treatment values analyzed by an ANCOVA, the sample size per group, the expected (or clinically relevant) effect, the assumed correlation between pre-treatment and post-treatment values, and the significance level are needed to calculate the power. In this simple case, the sample size and significance level are design choices and the rest are assumptions about the true nature of the treatment effect, while the only operating characteristic we are interested in is the power.
For designs with adaptive elements such as group sequential and platform trial designs, typically many more design choices and assumptions need to be made, while at the same time evaluating additional operating characteristics. As an example, in the case of a group sequential design with one interim analysis and the option to stop for futility, additionally, the time point of the interim analysis and the futility stopping boundary are needed as design parameters, the accrual rate is needed as an additional assumption and another operating characteristic we might evaluate is the average duration of the trial and the probability to stop at interim. Due to their flexibility with respect to incorporating adaptive design features, platform trials require even more design choices and assumptions for the simulation setup, and many more operating characteristics are evaluated.
\\
In this subsection, we will first give a description of the general simulation settings with the design choices and assumptions considered.  Then we specify operating characteristics of interest.
R version 4.2.1 was used for the simulation and the code is publicly available on Github \cite{github}. \\
The main objective of the simulations was to investigate whether a platform trial offers more efficiency in terms of sample size or time compared to separate 2-arm trials in the context of phase II MDD trials. Such a sequence of 2-arm trials will therefore act as a reference for the platform trial. 

\subsubsection{Simulated trial designs}
For the simulations we consider a base design with some fixed parameters and some that vary over a range of options. All parameters that are varied are described in Table \ref{tab:sim_setup}.  
\begin{table}[!htbp]
\centering
\caption{Parameters required to be specified for the simulation study. They are classified as either design choices or assumptions made regarding the platform trajectory or treatment effects. Some values are fixed and some varied for different simulation scenarios. Values marked by $^{\ast}$ are presented in the online supplement.}
\label{tab:sim_setup} 
\begin{tabularx}{\textwidth}{>{\hsize=.6\hsize\linewidth=\hsize}X>{\hsize=.4\hsize\linewidth=\hsize}X>{\hsize=.7\hsize\linewidth=\hsize}X>{\hsize=2.3\hsize\linewidth=\hsize}X}
\hline\noalign{\smallskip}
\hline\noalign{\smallskip}
Name & Type & Investigated Values & Description \\
\noalign{\smallskip}\hline\noalign{\smallskip}
\hline\noalign{\smallskip}
Randomisation & Design choice & Simple$^{\ast}$,  \newline (Modified) Block & For the simple randomisation the randomisation probabilities are set to achieve the targeted allocation ratios. A modified block randomisation was implemented which combines traditional block randomisation with random adding of controls to allow for minimal block length (details see \ref{sec:block}) \\
\hline\noalign{\smallskip}
Allocation ratio & Design choice & 1:...:1 \newline 1:...:1:k \newline 1:...:1:$\sqrt{k}$ \newline 1:...:1:$\sqrt{k}$ with minimum allocation cap to control $^{\ast}$ & Patients are randomised to one of the treatment arms or to the control arm with the given ratio. The variable $k$ denotes number of concurrently enrolling treatment arms. \newline 
$^{\ast}$ Different possible values for the cap on minimal allocation to control are investigated in the online supplement. Here the results of the selected cap 35\% are shown. \\
\hline\noalign{\smallskip}
Analysis method & Design choice & one covariate \newline two covariates & The analysis is based on an ANCOVA which either adjusts only for the covariate baseline value or additionally for the time period.\\
\noalign{\smallskip}\hline
Timing of interim analyses & Design choice & no interim analysis, \newline 50\% & Timing of the interim analysis of a single treatment arm as proportion of the planned sample size for this treatment arm (counting observed outcomes). \\
\hline\noalign{\smallskip}
Futility stopping rule & Design choice & no futility stopping, \newline 0.2, 0.25, 0.3, 0.35, 0.4, 0.45, 0.5 & Treatment arms will be stopped for futility at the time point of the interim analysis if the one-sided interim p-value is above this threshold. \\
\hline\noalign{\smallskip}
Total sample size per treatment & Design choice & variable options from 40 to 120 & Number of patients after which the final analysis of one treatment arm is conducted. \\
\hline\noalign{\smallskip}
Initial treatments & Assumption & 3, \newline 6 & Number of treatments available at the beginning of the platform trial\\
\hline\noalign{\smallskip}
Timing of new treatments & Assumption & 20\%, \newline 100\% & Every month a new treatment can enter the platform with a given probability if the maximum number of concurrently enrolling treatment arms is not yet reached. \\
\noalign{\smallskip}\hline
Standardized effect size & Assumption & 0, \newline 0.2, \newline 0.35, \newline 0.5 & Every treatment that enters the platform trial is randomly assigned one of these standardized effect sizes $d$ with predefined probability $\theta_d$, with $\theta_0 + \theta_{0.2} + \theta_{0.35} + \theta_{0.5} = 1$. The effect sizes are expressed in standardized mean difference between baseline and 6-week MADRS score. A variance-covariance matrix based on data from past studies is used for standardization. \\
\noalign{\smallskip}\hline
Effect size \newline distribution & Assumption & equal, \newline pessimistic & Different scenarios for the probabilities of the effect sizes are investigated. The equal scenario sets $\theta_0 = \theta_{0.2} = \theta_{0.35} = \theta_{0.5} = 0.25$ and the pessimistic scenario $\theta_0 = 0.5$, $\theta_{0.2} = 0.3$, $\theta_{0.35} = 0.1$, and $\theta_{0.5} = 0.1$.\\
\noalign{\smallskip}\hline
Time trend & Assumption & 0\%, \newline 10\%  & The assumed drift over time is modelled by a step function with steps at the beginning of every time period. The step width is given as a percentage in terms of the 6-week MADRS variance. 
\end{tabularx}
\end{table}
As the base design, we used a platform trial with 6 concurrently running treatment arms and a common control group. The targeted sample size per treatment arm $j$ is $n_j=80$ patients and for the treatment arms, each of the four standardized effect sizes $d=0$, $d=0.2$, $d=0.35$, and $d=0.5$ are assumed as equally likely, i.e. the probability of each effect size is $\theta_0 = \theta_{0.2} = \theta_{0.35} = \theta_{0.5} = 0.25$.
Every week a mean number of 7 patients is recruited with 7 being most likely (90\%) and a slight variability to 6 or 8 patients per week (both with probability 5\%). Also every week a final analysis can take place if enough patients are allocated to an arm. In the simulations, we assume that the outcome is observed immediately and analyses (and possible trial adaptations) are conducted as soon as a target number of outcomes have been observed. Therefore, active arms and enrolling arms are synonymous in the simulations. The decisions are based on p-values using an ANCOVA adjusting for the baseline value and a nominal one-sided significance level $\alpha = 0.05$ was used.
\\
For each combination of simulation parameters, 10000 simulation runs were performed. 10000 replicates correspond to a simulation error for a rate of 0.05 (=significance level) of $\sqrt{0.05*(1-0.05)/10000} \approx 0.002$ and the worst case simulation error for rates of $\sqrt{(0.5*0.5)/10000}=0.005$.
\\
At the beginning of each month, if a treatment arm has been removed from the platform, another treatment arm replaces it. For simulation purposes, we set one month to 4 weeks. This scenario uses an optimistic assumption that there are always treatments available to be entered into the platform trial and shows the maximum benefit of a platform trial compared to multiple 2-arm trials.
\\
After 60 months, no more new treatment arms can be added and the platform ends once all compounds have made a decision. In order to archive a more homogeneous total sample size in the investigated platform trial designs, we only allow treatments to enter if at least one-fifth of the targeted sample size is expected to be recruited until month 60. The distribution of the platform sample size in case this rule is not applied is provided in the online supplement.
\\
We also investigated another scenario that makes more pessimistic and maybe more realistic assumptions about the availability of new treatment arms and the effect size distribution. 
In this second scenario, the platform starts with 3 arms and at the beginning of each month, a new arm enters the trial with 20\% probability if the maximum number of 6 concurrently enrolling treatments in the platform is not yet reached. In the comparison section to multiple 2-arm trials, results for this scenario are shown additionally to the results of the platform running at maximum capacity. More results for this setting are provided in the online supplement.
\\ 
For the effect size distribution, we investigate a more pessimistic scenario with probabilities of assignment $\theta_0 =0.5$, $\theta_{0.2} =0.3$, $\theta_{0.35} =0.1$, and $\theta_{0.5} = 0.1$. Results for the pessimistic scenario are shown in the futility section. Additional results for this setting are provided in the online supplement.
\\
Starting with this base design we investigate different design elements separately. We first evaluate the impact of different allocation ratios to control and select the most promising one which will be fixed for the following simulations. Secondly, we show how the power is impacted when adjusting for time periods in the final analysis to address potential changes in the treatment effect due to adding and dropping of treatment arms. Thirdly, an interim analysis is included in the platform trial after half of the targeted patients per treatment arm are accrued. We only consider early futility, which is triggered if the p-value is larger than a pre-defined futility boundary (different values for this futility boundary are investigated). 
In our design, we do not account for futility stopping in the significance level of the final analysis, therefore the futility boundaries can be considered non-binding. This offers more flexibility because they allow for overruling the stopping decision at interim. However, in order to present operating characteristics, predictability is needed. So in the visualization of the simulation results, we treat the boundaries as binding.
As a fourth step, we analyse the impact of the targeted sample size per treatment arm on the operating characteristics of the platform trial. 
\\
Additionally, in order to be able to compare the results of the platform trial to a more traditional approach, we simulated standard parallel group designs with 1:1 allocation between treatment and control both with and without an interim analysis after half of the targeted per-arm sample size is accrued. For comparability, we always use the same general assumptions (e.g. effect size and per-arm sample size) as for the platform trial design.
\subsubsection{Operating Characteristics} 
For the traditional parallel group design operating characteristics that are often determined are the power and type I error per investigated treatment, as well as the expected sample size. In the context of platform trials, many more operating characteristics can be of interest depending on the objective of the platform trial. In Table \ref{tab:ocs} we give an overview of operating characteristics we considered important for the platform trial in MDD.
As several treatments are investigated within a platform, operating characteristics can be defined on several levels, e.g. on the platform level or just for a subset of interest like treatment arms or a certain effect size. For example, one could be interested in the expected sample sizes both on the treatment and platform level for budgeting reasons.
\begin{table}[h!]
\centering
\caption{Operating characteristics of interest in platform trials. This table gives names and descriptions of different operating characteristics that will be evaluated in the simulation section. It also states on which levels the characteristics are calculated, i.e. if they are characteristics of the overall platform, of arms in the platform or stratified by different effect sizes.}
\label{tab:ocs} 
\begin{tabularx}{\textwidth}{>{\hsize=.7\hsize\linewidth=\hsize}X>{\hsize=.3\hsize\linewidth=\hsize}X>{\hsize=2\hsize\linewidth=\hsize}X}
\hline\noalign{\smallskip}
\hline\noalign{\smallskip}
Name & Level &  Description \\
\noalign{\smallskip}\hline\noalign{\smallskip}
\hline\noalign{\smallskip}
Rate of decisions made & per effect size & Decisions can be "success", "failure" or "stopped for futility". The proportion of decisions made is evaluated depending on the assumed effect size. If the effect size is $d = 0$ the success rate corresponds to the type I error. If the effect size is greater than $0$ the success rate corresponds to the power and the rates for failure and stopped for futility together give the type I error. \\
\noalign{\smallskip}\hline
Absolute number of rejected null hypotheses & platform, \newline per effect size & Gives the absolute number of treatments for which the null hypothesis can be rejected either within the whole platform or stratified by effect sizes. \\
\noalign{\smallskip}\hline
Sample size (median and interquartile range) & platform, \newline treatment & Gives the average number of patients in the whole platform trial, and the single treatments taking into account the potential interim analysis. \\
\noalign{\smallskip}\hline
Size of control group for interim analyses and final analyses (median and interquartile range) & platform, \newline treatment & Only concurrent controls are used for the interim and final analyses of a specific treatment. Because the allocation to control is not fixed, this value can vary notably. \\
\noalign{\smallskip}\hline
Number of treatment arms (median and interquartile range) & platform & Gives the average number of different treatments that can be investigated in the platform trial depending on the stopping criteria of the overall platform. \\
\noalign{\smallskip}\hline
Standardized expected number of treatment arms & platform & Gives the average number of different treatments that can be investigated per 1000 patients in the platform trial. \\
\noalign{\smallskip}\hline
Duration (median and interquartile range) & treatment & Gives the value of how long the platform will run and how long it takes on average for a treatment to finish within the platform trial. \\
\end{tabularx}
\end{table}
\section{Simulation Results}
We will first present the results of the impact of different allocation ratios to control. Secondly, we show how the power is influenced when adjusting for time periods in the analyses.
Thirdly, interim stopping rules are investigated followed by an analysis of the targeted sample size per treatment arm.
Finally, a platform trial for the selected allocation strategy and an interim analysis for futility is compared to a sequence of standalone 2-arm trials. 
\\
At the beginning of every subsection, we state the fixed parameters if they differ between the subsections. In this main paper, we only present a limited number of operating characteristics and simulation scenarios. We refer to the online supplement for more details. 
\subsection{Comparison of allocation ratios}\label{sec:alloc}
We simulated platform trials with different choices of allocation methods. The targeted sample size per treatment arm was set to $N=80$ and no interim analysis was conducted. 
The allocation methods considered are 1:1:...:1 randomisation (called balanced allocation), 1:1:...:$k$ randomisation (called $k$ allocation), 1:1:...:$\sqrt{k}$ (called $\sqrt{k}$ allocation) and 1:1:...:$\sqrt{k}$ with a cap at 35\% for control.
\begin{figure}[h!]
  \centering
  \includegraphics[width=17cm]{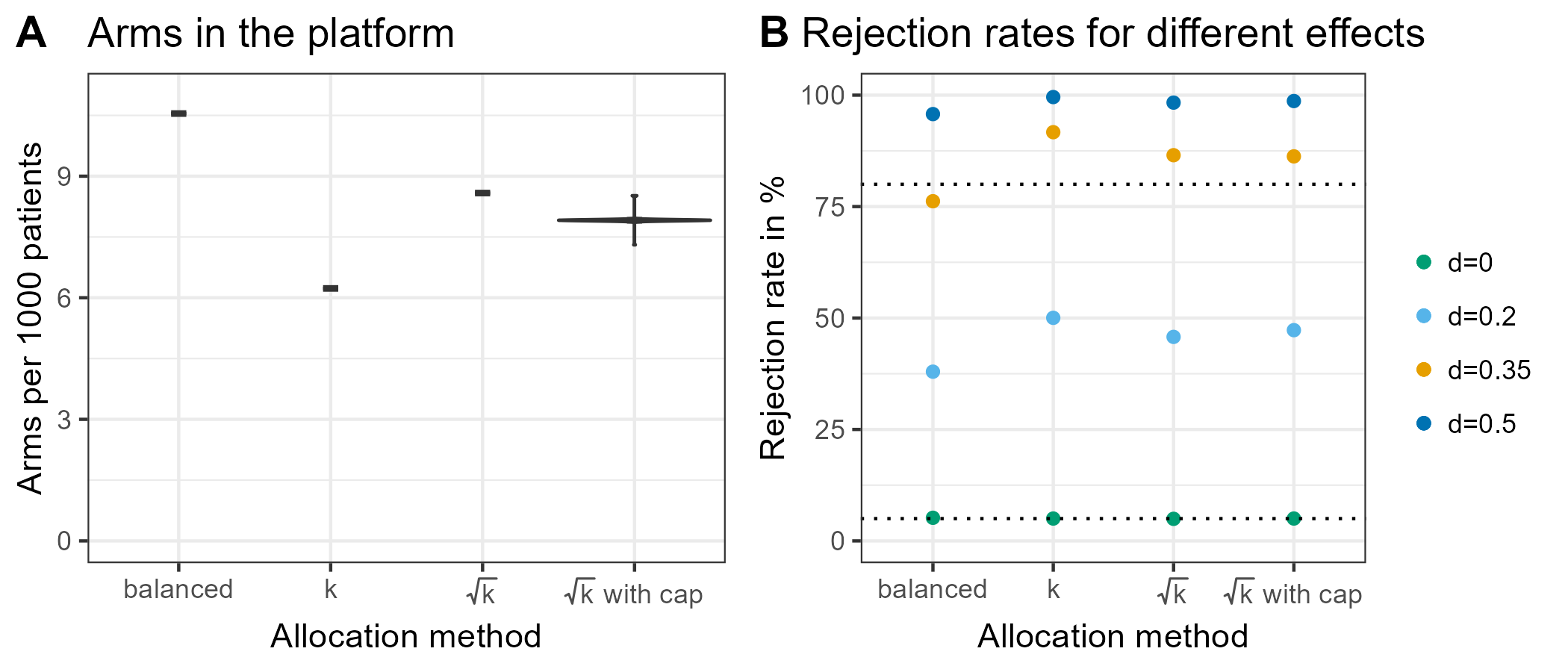}
  \caption{Treatment arms per 1000 patients in the platform, and rejection rates for platforms running at maximum capacity, i.e. always investigating 6 arms in parallel. The targeted sample size per treatment arm was fixed at $N=80$, and all effect sizes were assumed to be equally likely. The minimal control cap for the $\sqrt{k}$ allocation was set to 35\%. A) depicts the number of arms that can be evaluated per 1000 patients in a corresponding platform trial. In B the percentage of rejected null hypotheses is depicted. It equals the type I error rate for $d = 0$ and the power for the other values of $d$. The type I error rate is always controlled at 5\%. This value is indicated by the lower dotted line. The higher dotted line highlights the 80\% mark.}
    \label{fig:alloc}
\end{figure}
The median number of experimental treatment arms per 1000 patients in the platform trial is the highest for the balanced allocation with 10.5 arms. When applying the k allocation substantially fewer arms (6.2) can be investigated due to the higher proportion of controls. The $\sqrt{k}$ allocation and $\sqrt{k}$ allocation with a minimum allocation to control of 35\% are somewhere in between with a median of 8.6 and 7.9 arms per platform, as seen in Figure \ref{fig:alloc}A.
Directly correlated to the number of controls per comparison, the duration of experimental treatment arms is the shortest for the balanced 1:...:1 randomisation and the longest for the $k$ allocation. The arms in the scenario with $\sqrt{k}$ allocation take more time than the ones in the balanced allocation scenario and when using a minimum cap for control at 35\% they again take a little longer. 
\\
Figure \ref{fig:alloc}B shows the percentage of rejected null hypotheses in platform trials using the different allocation methods stratified by the four different standardized effect sizes $d=0$, $d=0.2$, $d=0.35$, and $d=0.5$. 
All effect sizes are assumed to be equally likely, i.e. $\theta_0 = \theta_{0.2} = \theta_{0.35} = \theta_{0.5} = 0.25$. For $d=0$ the rejection rate equals the type one error rate, which is always controlled at 5\%. In the other scenarios the power is always higher the higher the number of control comparators per decision. So the $k$ allocation yields the highest power followed by $\sqrt{k}$ allocation with a lower cap for control and $\sqrt{k}$ allocation without a cap. The lowest power is achieved using balanced allocation. It even yields under 80\% power in the case of $d=0.35$ (the minimal clinically relevant effect size) where all other methods have power clearly over 80\% power.
\\
Overall, even though the $k$ allocation method yields the highest power it also has the lowest ratio of patients on treatment and by far the highest duration per experimental treatment arm as well as the lowest (standardized) number of arms which can be investigated in the platform. The balanced allocation enables investigation of the most treatment arms per 1000 patients and the lowest ratio of patients on control but the power is much smaller than for the other allocation methods. As a trade-off between number of arms investigated and power, we selected the $\sqrt{k}$ allocation method with a minimum cap of 35\% for control. It achieves the second highest power of all investigated allocation methods and enables the investigation of more treatment arms within the platform than the balanced allocation method. Furthermore, this boundary was chosen based on clinical considerations: In patients with MDD, the placebo response fluctuates greatly with personal expectations of whether placebo is given or an investigational drug. The expectation is that the lower the probability of being randomised to the control arm, the greater the placebo response \cite{papakostas}. Therefore, clinicians advise randomising at least one-third to the control arm. A more thorough investigation of the impact of different values for the cap can be found in the online supplement.
\subsection{Comparison of ANCOVA models}
An ANCOVA is used to test the null hypothesis of no treatment effect against the one-sided alternative hypothesis. Potential covariates to include are the baseline value and the time period. Time period refers to a section of time during which no treatment leaves the trial and no new treatment enters. 
When assuming no time trend, additionally including the time period in the analysis leads to a very slight power loss, see Table \ref{tab:time_period}. However, one can never be sure if there really is no time trend in real-world settings. Table \ref{tab:time_period} illustrates the possible difference between including and not including the time period when a time trend is present.
\begin{table}
\centering
\caption{Power of ANCOVA models with and without adjustment for time periods as factor assuming no time trend and a time trend modeled by a step function assuming a step width of 10\% of the 6-weeks MADRS score variance. The power is stratified by different effect sizes.}
\label{tab:time_period} 
\begin{tabularx}{\textwidth}{XXXXX}
\hline\noalign{\smallskip}
\hline\noalign{\smallskip}
 & \multicolumn{2}{c}{no time trend assumed} & \multicolumn{2}{c}{stepwise time trend assumed} \\
Effect Size $d$ & without time period & with time period &  without time period & with time period \\
\noalign{\smallskip}\hline\noalign{\smallskip}
\hline\noalign{\smallskip}
0 & 4.97 & 4.95 & 4.99 & 4.98 \\
\noalign{\smallskip}\hline
0.2 & 48.3 & 47.9 & 46.2 & 48.1 \\
\noalign{\smallskip}\hline
0.35 & 87.9 & 87.7 & 86.7 & 88.2 \\
\noalign{\smallskip}\hline
0.5 & 99 & 98.9 & 98.8 & 99.1\\
\noalign{\smallskip}\hline
\end{tabularx}
\end{table}
The trend in this exemplary simulation is assumed to be a step function on the 6-week MADRS score with steps at time points where treatments enter or leave the platform trial (i.e. at the beginning of every time period). The step width is set to an increase of about 10\% of the variance of the 6-week MADRS score. This value and also the time trend model are exchangeable and only serve the purpose of illustrating the improvement of power in settings with time trends. 
As there is only a minimal loss in power when there is no time trend at all and a noticeable increase in power depending on the actual time trend, it is recommended to include the time period as a covariate in the ANCOVA for real world applications. However, our further simulations do not include assumptions on time trends and we therefore only used the baseline value as a covariate in the ANCOVA.
\subsection{Comparison of futility stopping rules}
We simulated platform trials with different choices of futility boundaries. We applied a $\sqrt{k}$ allocation with a minimal control cap at 35\% and a targeted sample size of 80 patients per experimental treatment arm. The interim analysis was carried out after 40 patients were accrued in a treatment arm. We present two different scenarios for the assignment probabilities of the four considered standardized effect sizes, an equal scenario and a more pessimistic one. In the equal scenario all effect sizes are assumed to be equally likely, i.e. $\theta_0 = \theta_{0.2} = \theta_{0.35} = \theta_{0.5} = 0.25$. For the more pessimistic scenario these probabilities of assignment are $\theta_0 =0.5$, $\theta_{0.2} =0.3$, $\theta_{0.35} =0.1$, and $\theta_{0.5} = 0.1$. In the online supplement we additionally report the observed probabilities for stopping for futility depending on the effect sizes.
\begin{figure}[h!]
  \centering
  \includegraphics[width=16cm]{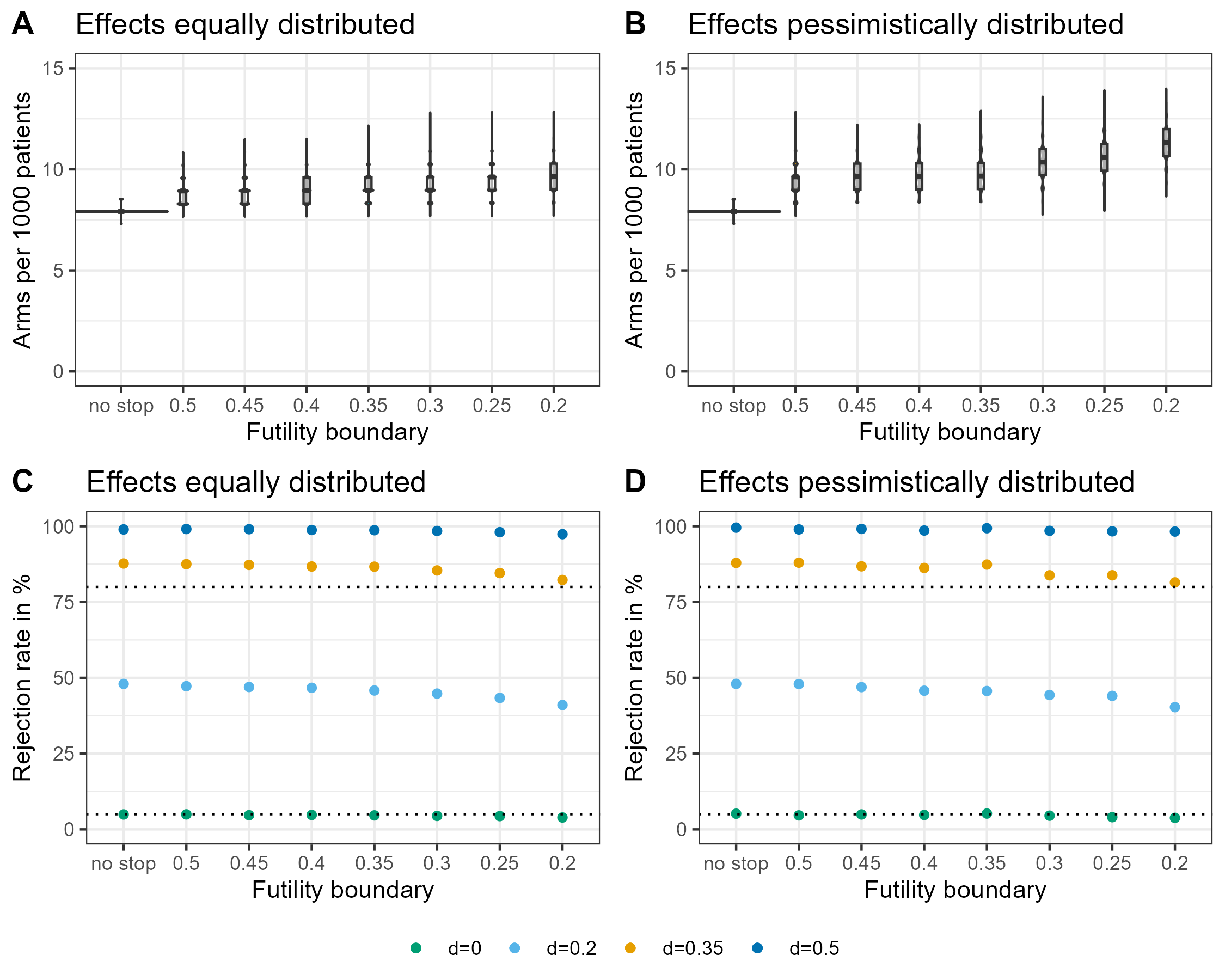}
    \caption{Rejection rates and standardized number of arms when implementing different futility rules. In the scenario on the left side (A and C) all effect sizes are assumed to be equally likely, i.e. $\theta_0 = \theta_{0.2} = \theta_{0.35} = \theta_{0.5} = 0.25$. On the right side (B and D) we present results for a more pessimistic scenario with probabilities of assignment $\theta_0 =0.5$, $\theta_{0.2} =0.3$, $\theta_{0.35} =0.1$, and $\theta_{0.5} = 0.1$. A and B give the standardized number of arms per 1000 patients in the platform trial. All arms are included in this number regardless which of the four different investigated effect sizes was allocated. In the more pessimistic scenario (B) more arms can be investigated. The difference becomes more prominent the stricter the futility boundary. In B and D the percentage of rejected null hypotheses is depicted. It equals the type I error rate for $d = 0$ and the power for the other values of $d$. The type I error rate is always controlled at 5\%. This value is indicated by the lower dotted line. The higher dotted line highlights the 80\% mark. The power decreases with stricter futility boundaries. Overall C and D are quite similar. So the effect size scenario does not have a big impact on the rejection rates.}
    \label{fig:fut_comp}
\end{figure}
Overall the rejection rates, seen in Figure \ref{fig:fut_comp}C and D are quite similar in both scenarios. With a more strict stopping rule, the duration of the platform decreases as is the median number of control comparators per decision. As a trade-off for using non-binding futility boundaries, the power decreases with stricter boundaries but also more arms can be tested. The number of arms that can be tested per 1000 patients in the platform differs notably between the scenarios, see Figure \ref{fig:fut_comp}A and B. In the more pessimistic scenario more arms can be investigated than in the equal scenario. This difference becomes more prominent the stricter the applied futility boundary and goes as high as 1.45 arms per 1000 patients for a futility boundary of 0.2. Here we only present the overall number of arms. 
In the equal scenario the same number of arms can be investigated for each effect size. In the more pessimistic scenario the most arms can be investigated for $d=0$ followed by $d=0.2$ and then tied for the last place $d=0.35$ and $d=0.5$, corresponding to the probability of occurrence of the individual effect sizes.
Figure \ref{fig:fut_comp} shows that in both scenarios the difference between not stopping at all and a very soft boundary of 0.5 is relatively large for the standardized number of arms that can be evaluated in the platform. The impact on the power, however, is very small. The differences in these operating characteristics between the single steps for the futility boundary are not very pronounced when comparing one step with the next. 
However, the differences become clearly perceptible when values further apart are compared, like the boundaries 0.5 and 0.25. Here, however, not only does the number of arms that can be examined increase, but the power also decreases notably. So one has to decide for the special use case how aggressive the applied stopping rules should be. As this is a phase II study, we recommend applying futility stopping with a boundary of at least 0.5. It is also quite an intuitive boundary because it stops the investigation when the treatment effect points in the opposite direction.
\subsection{Analysis of per arm sample sizes}
\begin{figure}[h!]
  \centering
  \includegraphics[width=16cm]{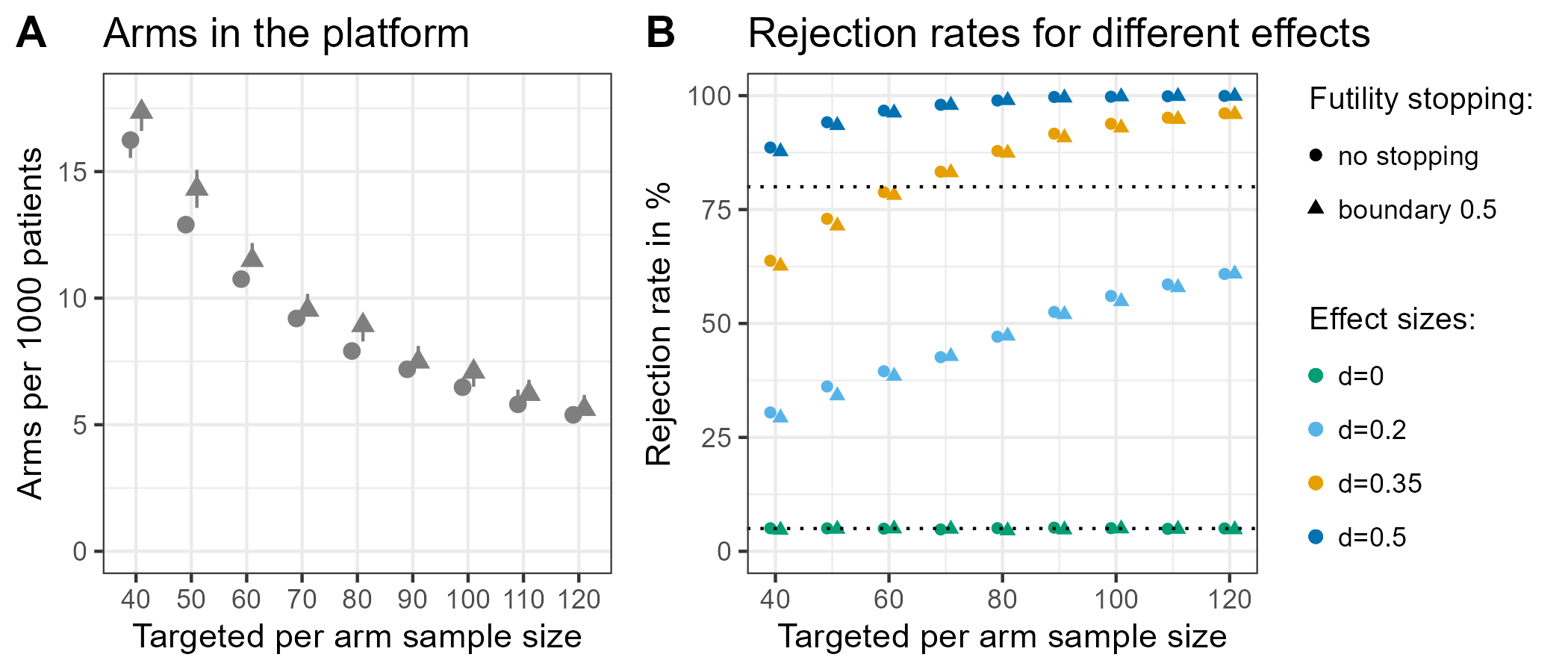}
    \caption{Rejection rates and standardized number of arms for different per-arm sample sizes. All effect sizes are assumed to be equally likely, i.e. $\theta_0 = \theta_{0.2} = \theta_{0.35} = \theta_{0.5} = 0.25$. The circles give the values without the implementation of an interim analysis and the triangles the corresponding values when a futility boundary of 0.5 is applied. The sample size depicted on the x-axis was examined in steps of 10. The small variation in x direction is based on jittering for better readability. Figure A) shows the median number of arms per 1000 patients and the interquartile range. In B) the rejection rates are depicted stratified by the four different investigated effect sizes. The rejection rate equals the type I error rate for $d=0$ and the power for the other values of $d$. The type I error rate is always controlled at 5\%. This value is indicated by the lower dotted line. The higher dotted line highlights the 80\% mark. The power is higher the higher the targeted sample size but fewer arms can be investigated. When futility stopping is implemented generally more arms can be investigated and the power is lower.}
    \label{fig:n_arms_power}
\end{figure}

We simulated platform trials with different choices of the per treatment arm sample size. The $\sqrt{k}$ allocation with a minimal control cap of 35\% was used and we investigated designs without the option to stop for futility and with an interim analysis after 50\% of the targeted sample size and a futility boundary of 0.5.
The higher the targeted sample size per treatment arm the longer the investigation of this treatment arm takes but the higher the power. Figure \ref{fig:n_arms_power}B shows the power reached when targeting different per treatment arm sample sizes spanning from $N=40$ to $N=120$ with a step width of 10. The standardized number of arms that can be investigated is depicted in \ref{fig:n_arms_power}A.  
The values with inclusion of the interim analysis are depicted by triangles and the values without the option to stop for futility are depicted by circles.
Applying the specified interim analysis, for the minimal clinical relevant effect size $d=0.35$ a sample size of $N=60$ gives a power of 78.2\%. It therefore falls a little short when wanting to reach 80\%. When the futility boundary 0.5 is used $N=70$ is the smallest investigated sample size with power above 80\% and $N=100$ the smallest with power above 90\%. It is most cost-efficient to use the smallest sufficient sample size for the desired power. But the problem with a smaller sample size is that it introduces higher variability and it becomes more likely to stop for futility. In the next subsection, we therefore use the sample sizes between 60 and 90 for the comparison between platform trials and traditional randomised controlled trials, again with a step-width of 10.
\subsection{Comparison to Traditional Parallel Group Design}\label{sec:comparison}
\begin{figure}[h!]
  \centering
  \includegraphics[width=16cm]{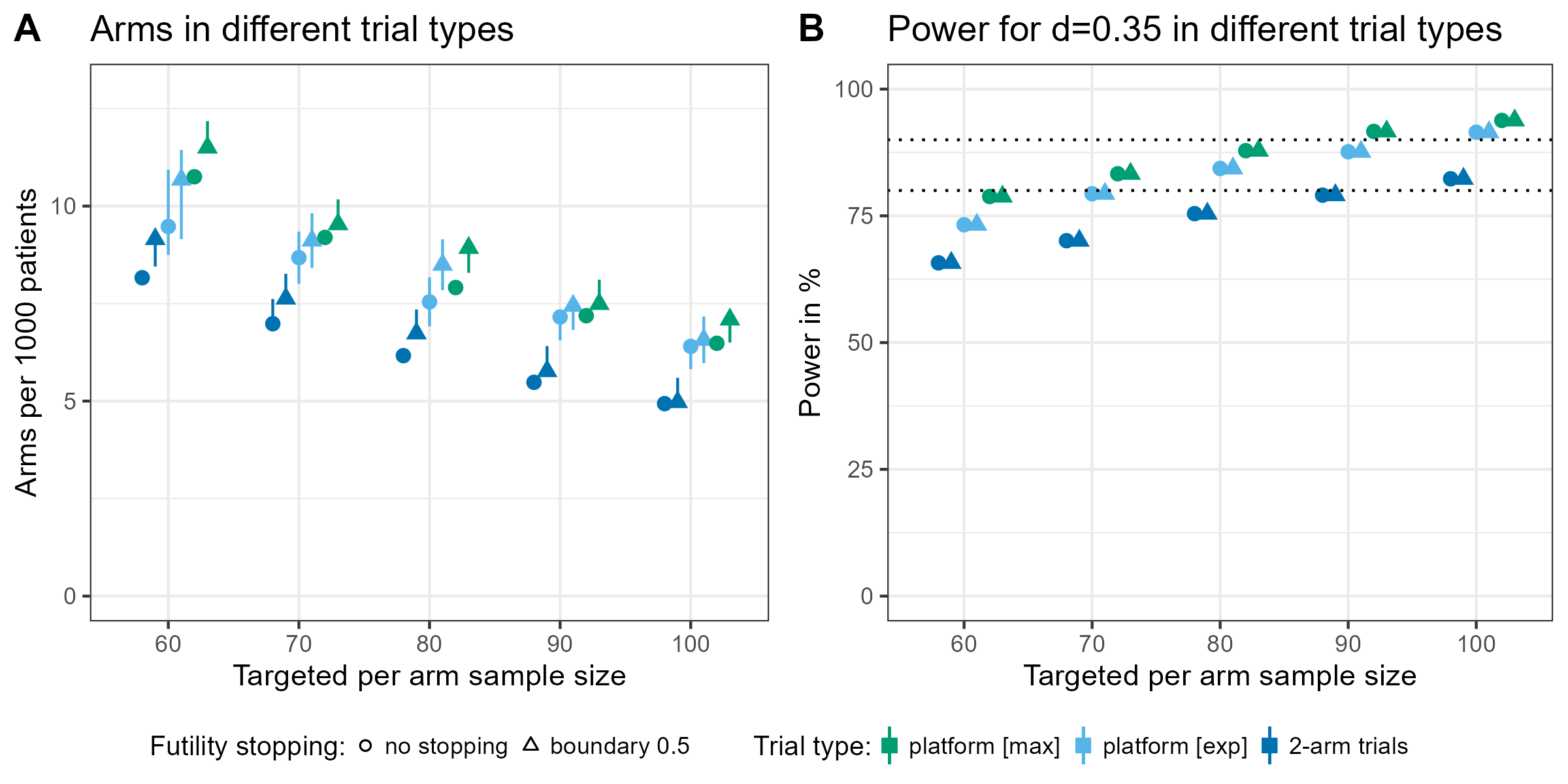}
    \caption{Comparison of operating characteristics in different trial types. All effect sizes are assumed to be equally likely, i.e. $\theta_0 = \theta_{0.2} = \theta_{0.35} = \theta_{0.5} = 0.25$. The circles give the values without the implementation of an interim analysis and the triangles the corresponding values when a futility boundary of 0.5 is applied. The sample size depicted on the x-axis was examined in steps of 10. The small variation in x direction is based on jittering for better readability. Figure A) shows the median number of arms per 1000 patients and the interquartile range for the three different trial types platform trial with maximum capacity utilization, platform trial with expected load in the MDD case, and the traditional approach with a series of individual 2-arm randomised controlled trials. B) gives the power for the same type of trials. A series of traditional 2-arm trials reaches the lowest power and the fewest arms can be investigated per 1000 patients. Incorporation of the possibility to stop for futility always results in evaluation of more arms and a slightly smaller power.
      }
    \label{fig:rct_comp}
\end{figure}

The classical approach for evaluating multiple treatments in the context of one disease, especially if more than one company is involved, is using multiple randomised controlled trial designs with one treatment arm and one control arm and 1:1 allocation between treatment and control. 
We simulated a series of such traditional trials in order to compare the operating characteristics to those of the designed platform trial. For the 2-arm trials we make the same assumptions regarding sample size per treatment arm, assumed effect sizes (and their distribution), and analysis method as in the platform design. We also included the option to stop for futility at an interim analysis after half of the total sample size is accrued.
Figure \ref{fig:rct_comp} shows the results of the comparison between sequential 2-arm trials and platform trials. The circles give the values for designs without interim analysis for futility and the triangles the values with a futility analysis after 50\% of the sample size was reached and with a futility boundary of 0.5.
Besides the platform trial running at maximum capacity, i.e. always running 6 treatment arms in parallel, we also included a more realistic scenario in the comparison. This is important as a platform running at maximum capacity shows the maximum benefit reached when using a platform trial but this maximum benefit is often not achieved in practice. The more realistic scenario considers the platform to start with three treatment arms in parallel and at the beginning of every month the probability of a new treatment arm to enter the platform is 20\%.
Figure \ref{fig:rct_comp}A shows the number of arms that can be investigated per 1000 patients. The inclusion of a futility analysis leads in all cases to a larger number of arms that can be investigated per 1000 patients. The platform trial running at maximum capacity enables evaluation of the most arms. However, the difference to the more realistic platform setting diminishes with bigger targeted sample sizes per treatment arm. This is due to the treatment arms staying longer in the platform if more patients are needed and therefore often more arms run in parallel. An important benefit of using platform trials is generated by investigation of multiple treatment arms in parallel. By sharing of the control group fewer patients per treatment arm are needed in the control group and the number of control comparators increases and therefore also the power. This, however, only holds true for some allocation methods like the $\sqrt{k}$ allocation method with a minimum cap for control used here. If a 1:1:...:1 allocation was to be applied in the platform the power would not increase as much and one could just use a traditional 2-arm trial with 1:1 allocation.
With the platform design options used in the simulations we always get higher power than for a series of 2-arm trials even when the platform does not run at full capacity, see Figure \ref{fig:rct_comp}B. The figure only shows the power for the minimal clinically relevant effect size $d=0.35$ because this effect is the most relevant when deciding about the sample size and design of a trial. When using 2-arm trials, only with a per-arm sample size of 100 we reach a power of over 80\% for this minimal clinically relevant effect size. In a platform running at maximum capacity a per-arm sample size of 70 suffices and in the more realistic workload setting a per-arm sample size of 80 would be needed. In a real world setting we cannot be sure about the actual workload the platform trial will have. Hence, the decision about the targeted per-arm sample size should not be made solely based on the power reached in a platform trial running at maximum capacity. For our specific use-case we would instead recommend a targeted per-arm sample size of 80 when a power of 80\% is desired. 
\subsection{Summary and recommendations}
Based on our simulations we recommend the use of $\sqrt{k}$ allocation with a minimal cap for control at 35\%, inclusion of time period as a factor (even though we do not include it in further simulations) and futility stopping with at least a boundary of 0.5, or even stricter like e.g. 0.25 if the platform is used for screening purposes. Like the futility boundary the sample size also depends on the specific use case. It is especially influenced by the desired power. Note that the power reached is also dependent on the available concurrent treatment arms, as we show in section \ref{sec:comparison}. One should therefore select a sample size that is a little higher than the minimum needed for the desired power. 
\\
Overall, the implementation of a platform trial provides both a higher power and a larger number of arms that can be tested with the same amount of patients. It should therefore be preferred in settings when it is anticipated that multiple treatments would participate. If one is unsure about the potential of the availability of new treatments to be investigated in the specific disease context and only very few are anticipated, the traditional 2-arm trial approach should be preferred. The organizational effort involved in launching a platform would then simply not be justified by the gain.
\section{Discussion}
This paper summarizes the design considerations made by an EU-PEARL working group for a phase II platform trial in MDD \cite{eu-pearl-mdd}, including EMA consultation at an ITF meeting. The simulations were conducted to evaluate if running a platform trial would be feasible especially compared to separate 2-arm trials. The simulations were also needed to fine-tune the design elements, such as decision rules, analysis strategy, randomisation method, and sample size. Due to the many random influences inherent in platform studies, it is not possible to calculate the operating characteristics deterministically.
\\
The use of platform trials allows great flexibility and individual design elements such as futility bounds can be individually selected to best suit the specific application. For reproducibility, we made our extensive simulation code available on Github \cite{github}. An overview of different available simulation software for platform trials can be found in Meyer et al. \cite{meyer_software}. However, the code for the simulation studies is very individual, as are the platform studies themselves. Often it is hard to generalize code and large parts have to be rewritten. Recently, more generally applicable simulation and visualisation tools for platform trials have been proposed \cite{meyer2023simple, meyer2023shiny}.
\\
Platform trials allow for great flexibility and many additional design elements might be included. 
For example, by using group sequential boundaries stopping for efficacy could be incorporated. Also, the platform might allow different arms to target different sample sizes per experimental treatment arm. This could, for example, allow arms of sponsors targeting 80\% power and arms of those targeting 90\% power to be accommodated in the same platform trial. However, the specific implications of this would need to be explored separately. Assumptions about frequency distributions of the different targeted per-arm sample sizes at different time points in the platform would have to be made in order to investigate how this changes the respective concurrent control arms and thus also the power. \\
Since most available treatments are administered orally, we focus the investigation on this way of administration although we discussed platform designs for MDD including multiple ways of administration in the process. Allowing different ways of administration (e.g. intravenous or intranasal) would require having multiple control arms because otherwise appropriate blinding could not be ensured. Having more control arms would however result in smaller allocation rates for individual treatment arms and thus also longer durations per investigation. Overall the benefit of using a platform trial would be split between the different ways of allocation. Because non-oral treatments are rather rare we therefore recommend investigating treatments with different ways of administration  in separate trials. \\
Additionally, in the context of MDD patients with treatment resistant depression (TRD) one could consider the possibility of allowing patients who had no treatment benefit to re-enter the platform trial after participation.  
In case of allowing different routes of administration, re-entry could be restricted to another way of administration.
However, the statistical implications would need to be further investigated.
\\
In this article, we have focused on an adaptive allocation method with the allocation ratio to control depending on the actual number of open treatment arms. More controversially discussed are adaptive allocation methods also utilizing information on the observed treatment effect, such as response adaptive randomisation \cite{thall2015statistical, wathen2017simulation, proschan2020resist, villar2021temptation, robertson2023response}. Challenges arise if considering response adaptive randomisation in the presence of time trends, as response adaptive randomisation could lead to biased estimates if the analyses used are not appropriate for this situation. The current literature reflects considerations in more traditional 2-arm \cite{korn2022time} and multi-arm designs \cite{villar2018response}, but not in (perpetual) platform trials, where more information for estimation of potential time trends is available and/or the goal of response adaptive randomisation might only be to accelerate evaluation of beneficial arms, while all arms reach their targeted sample size eventually. Investigation of how response adaptive randomisation would affect the efficiency of platform trials and the performance of their analyses remains open for future work.
\\
There has been some debate about whether multiplicity adjustment is required when comparing multiple treatment arms with a common control in a platform trial without a consensus being reached \cite{molloy2021, nguyen_multi}. Guidance documents of EU- and US-authorities do not give a clear answer on when or how adjustment for multiplicity should be included in platform trials \cite{ema_cct, fda_cct, fda_master}. If separate trials were run no adjustment for multiplicity would be required and e.g. in platform trials with individual control arms, the correlation of test statistics leads to the family-wise error rate being lower compared to running individual trials with individual controls \cite{nguyen_multi}. Therefore on the one hand, it could be argued that no adjustment for multiplicity should be required in a platform trial setting. On the other hand, the dependency in the test statistics impacts the decisions made and should be considered especially in confirmatory settings \cite{collignon2020}. Another point of discussion is the interdependence of the individual hypotheses. At the moment the general approach is that no adjustment for multiplicity has to take place when the hypotheses are fairly unrelated, e.g. the treatments come from different sponsors and use different mechanisms of action. However, when the hypotheses are related, e.g. when several doses of one drug are tested, adjustment should take place \cite{wason2016, hungwang}. In our platform trial we will not adjust for multiplicity as we design an exploratory phase II trial and the hypotheses tested can be considered inferentially independent.
\\
Overall, based on our simulations we recommend a phase II platform trial in MDD using $\sqrt{k}$ allocation with a minimal control cap of 35\%, futility analyses after 50\% of the targeted per treatment arm sample size with futility boundary 0.5, inclusion of time period as a covariate in the ANCOVA analysis and a targeted per treatment arm sample size of 80 patients.
However, one should carefully investigate the different design choices for the specific use-case as they impact the performance characteristics greatly. 
\\
We have shown that a platform trial offers a more efficient way to test more treatments compared to a sequence of separate 2-arm trials each with its own control arm and they also achieve greater power in the evaluation of individual arms. 
The benefit depends on how many treatments are actually enrolling concurrently. The more enrolling treatments, the greater the benefit as less control data is required compared to the traditional approach. To allow the testing of more treatments, interim futility analyses should be performed to eliminate treatments that have either no or negligible treatment effects. 
\section*{Funding}
EU-PEARL (EU Patient-cEntric clinicAl tRial pLatforms) project has received funding from the Innovative Medicines Initiative (IMI) 2 Joint Undertaking (JU) under grant agreement No 853966. This Joint Undertaking receives support from the European Union’s Horizon 2020 research and innovation programme and EFPIA and Children’s Tumor Foundation, Global Alliance for TB Drug Development non-profit organisation, Springworks Therapeutics Inc. This publication reflects the authors’ views. Neither IMI nor the European Union, EFPIA, or any Associated Partners are responsible for any use that may be made of the information contained herein.  \\
MMF additionally received funding from the German Research Foundation (Project number \mbox{RA 2347/11-1}).
\section*{Acknowledgements}
The authors are grateful to the EU-PEARL investigators who contributed to the development of the MDD master protocol. 
The EU-PEARL MDD investigators are: 
Jelena Brasanac, Woo Ri Chae, Michaela Maria Freitag, Stefan Gold, Eugenia Kulakova, Christian Otte, Dario Zocholl (Charité - Universitätsmedizin Berlin), 
Marta Bofill-Roig, Elias Laurin Meyer, Franz König, Martin Posch (Medizinische Universität Wien), 
Heidi de Smedt, Yanina Flossbach (Janssen Pharmaceutica NV),
Melissa Kose, Giulia Lombardi, Carmine Pariante, Luca Sforzini, Courtney Worrell (King’s College London),
Tasneem Arsiwala, Alexandra Bobirca (Novartis Pharma AG),
Fernanda Baroso de Sousa, Pol Ibanez-Jimenez, Gabriela Perez-Fuentes, Toni Ramos-Quiroga (Fundació Hospital Universitari Vall d´Hebron – Institut de Recerca),
Witte Hoogendijk (Erasmus Universitair Medisch Centrum Rotterdam), 
Francesco Benedetti (Universita Vita-Salute San Raffaele), 
Fanni Laura Mäntylä (GAMIAN-Europe) 
\section*{Author contributions}
Conceptualization: MMF, DZ, SMG, MP, FK \\ 
Methodology: MMF, DZ, MP, FK \\ 
Software: MMF, DZ, ELM \\ 
Conduct of simulation study: MMF \\ 
Supervision: SG, MP, FK \\ 
Validation: FK \\ 
Visualization: MMF \\ 
Writing – original draft: MMF, FK \\ 
Writing – review \& editing: MMF, DZ, ELM, SMG, MBR, HDS, MP, FK 
\section*{Conflict of interest}
ELM is a salaried employee of Berry Consultants.
SMG reports honoraria from Hexal and Streamed-up.
All other authors did not report any conflict of interest.
\section*{Data availability statement}
The R code is made publicly available on Github \cite{github}.
\section*{ORCID}
Michaela Maria Freitag https://orcid.org/0009-0009-0924-1277\\
Dario Zocholl https://orcid.org/0000-0002-9218-6919 \\
Elias Laurin Meyer https://orcid.org/0000-0001-5398-6334\\
Stefan M. Gold https://orcid.org/0000-0001-5188-4799 \\
Marta Bofill Roig https://orcid.org/0000-0002-4400-7541\\
Martin Posch https://orcid.org/0000-0001-8499-8573 \\
Franz König https://orcid.org/0000-0002-6893-3304
\section{Bibliography}
\printbibliography
\end{document}


\maketitle

\section*{Funding}

EU-PEARL (EU Patient-cEntric clinicAl tRial pLatforms) project has received funding from the Innovative Medicines Initiative (IMI) 2 Joint Undertaking (JU) under grant agreement No 853966. This Joint Undertaking receives support from the European Union’s Horizon 2020 research and innovation programme and EFPIA and Children’s Tumor Foundation, Global Alliance for TB Drug Development non-profit organisation, Springworks Therapeutics Inc. This publication reflects the authors’ views. Neither IMI nor the European Union, EFPIA, or any Associated Partners are responsible for any use that may be made of the information contained herein.  \\
MMF additionally received funding from the German Research Foundation (Project number RA 2347/11-1).

\clearpage

\section{Comparison of simple randomisation to (modified) block randomisation}
We exemplary compare the impact of simple randomisation vs modified block randomisation for the $\sqrt{k}$ allocation with a cap for control in the basic platform design. The sample size is $n_j= 80$ patients per treatment arm $j$, no interim analysis takes place and the equal effect scenario is assumed ($d$=0, 0.2, 0.35, and 0.5 each with an a-priori probability of 25\%). The platform is assumed to be running at maximum capacity (with $k=6$ concurrently running experimental treatment arms), see main paper for more details on the comparison setting for allocation methods.
\begin{figure}[h!]
  \centering
  \includegraphics[width=16cm]{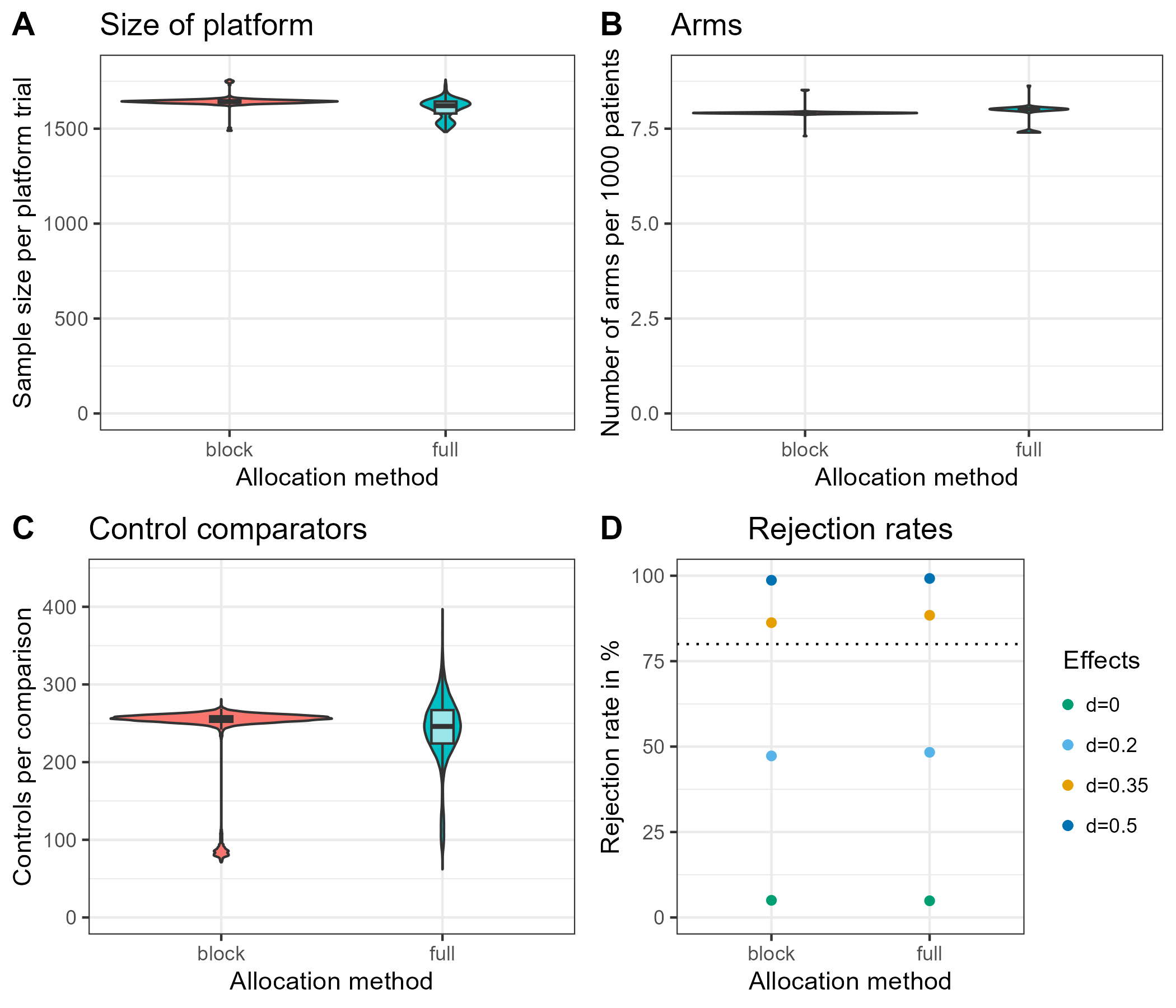}
    \caption{Sample size, number of arms, control comparators, and rejection rates per platform trial. All effect sizes are assumed to be equally likely, i.e. the probability of each effect size $\theta_0 = \theta_{0.2} = \theta_{0.35} = \theta_{0.5} = 0.25$. A) depicts the sample size of the platform trial and B) depicts the number of arms that can be evaluated in a corresponding platform trial standardized per 1000 patients in the platform. In C the number of control patients used per analysis is shown. The figures use violin plots with integrated box plots. The median sample size and the median number of arms are quite similar with a higher variability in case of simple randomisation. D) shows the rejection rates for different effect sizes, all being equally likely to occur in the platform.}
    \label{fig:full_block}
\end{figure}
The 1:$\sqrt{k}$ randomisation with minimum allocation to control of 35\% has a median platform sample size of about 1640 for the block randomisation and 1620 for the simple randomisation with the simple randomisation having a much higher variability, especially towards lower sample sizes, see Figure \ref{fig:full_block}A. The median number of arms is also quite similar and again there is a higher variability for the simple randomisation. Figure \ref{fig:full_block}B shows the number of arms that can be investigated per 1000 patients in a platform trial. The standardization was performed to make it more comparable. The higher variability especially towards lower numbers occurs because due to random effects, many patients can be randomised to the control arm and therefore fewer patients get randomised in treatment arms.
The additional variability introduced by using simple randomisation is especially prominent in the number of controls used per comparison, see Figure \ref{fig:full_block}C. In Figure \ref{fig:full_block}D The rejection rates for the different effect sizes are shown. The rejection rate corresponds to the type I error for $d=0$ and in the other cases to the power. The power is sometimes a little higher for the simple randomisation. This is might be related to the possibility of many more patients being actually randomised to the control arm due to random fluctuations by using the simple randomisation process.

\clearpage

\section{Comparison of platform end criteria}
\begin{figure}[h!]
  \centering
  \includegraphics[width=16cm]{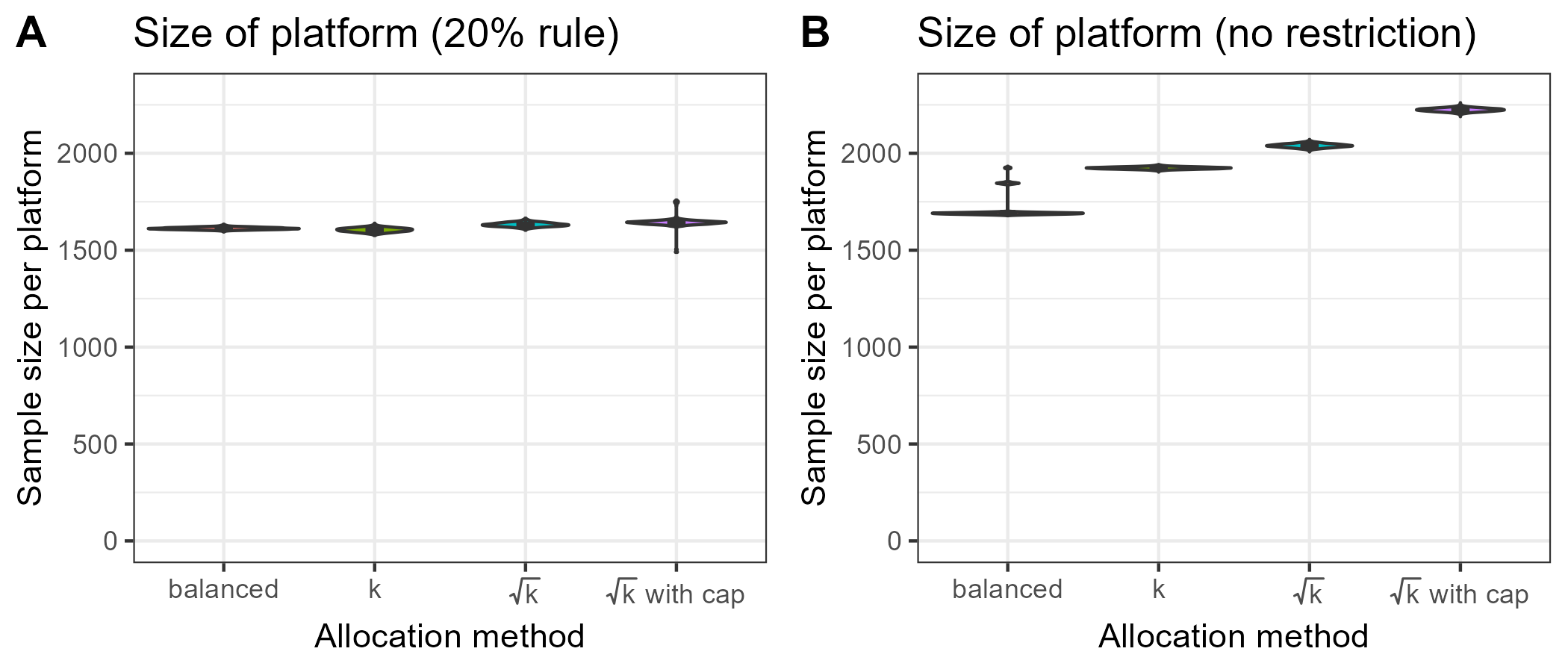}
    \caption{Comparison of different end criteria for the platform trial. A) shows the overall sample size of the platform trial for different allocation methods if treatment arms can enter the platform only if at least 20\% of the desired per-arm sample size can be accrued up to month 60. B) shows the corresponding values if treatment arms can enter the platform up to month 60 with no additional rule applied. For both the per-arm sample size was set to 80, no interim analyses were conducted and an equal effect size distribution was assumed, i.e. the probability of each effect size $\theta_0 = \theta_{0.2} = \theta_{0.35} = \theta_{0.5} = 0.25$.}
    \label{fig:end_rule}
\end{figure}

Platform trials can potentially run perpetually if no end criterion is specified. For the platform trial in MDD, the idea is to enable new treatment arms to enter the platform up to month 60. After this time point all arms still in the trial continue enrolling patients until a decision is made. This rule leads quite varying overall sample sizes in the platform trial if different design options are applied, like different allocation methods, see Figure \ref{fig:end_rule}B. 
In order to facilitate better comparability between different design options an additional condition was introduced in the simulation study that must be fulfilled for new treatment arms to enter the platform trial. They can only enter if it is expected that at least 20\% of the desired per arm sample size can be accrued up till month 60. This rule leads to a much more homogeneous overall sample size of the platform trial. Figure \ref{fig:end_rule} shows the results exemplary for the different allocation methods.
\vspace{1.5cm}

\clearpage

\section{Selection of the minimal control cap}
\begin{figure}[h!]
  \centering
  \includegraphics[width=16cm]{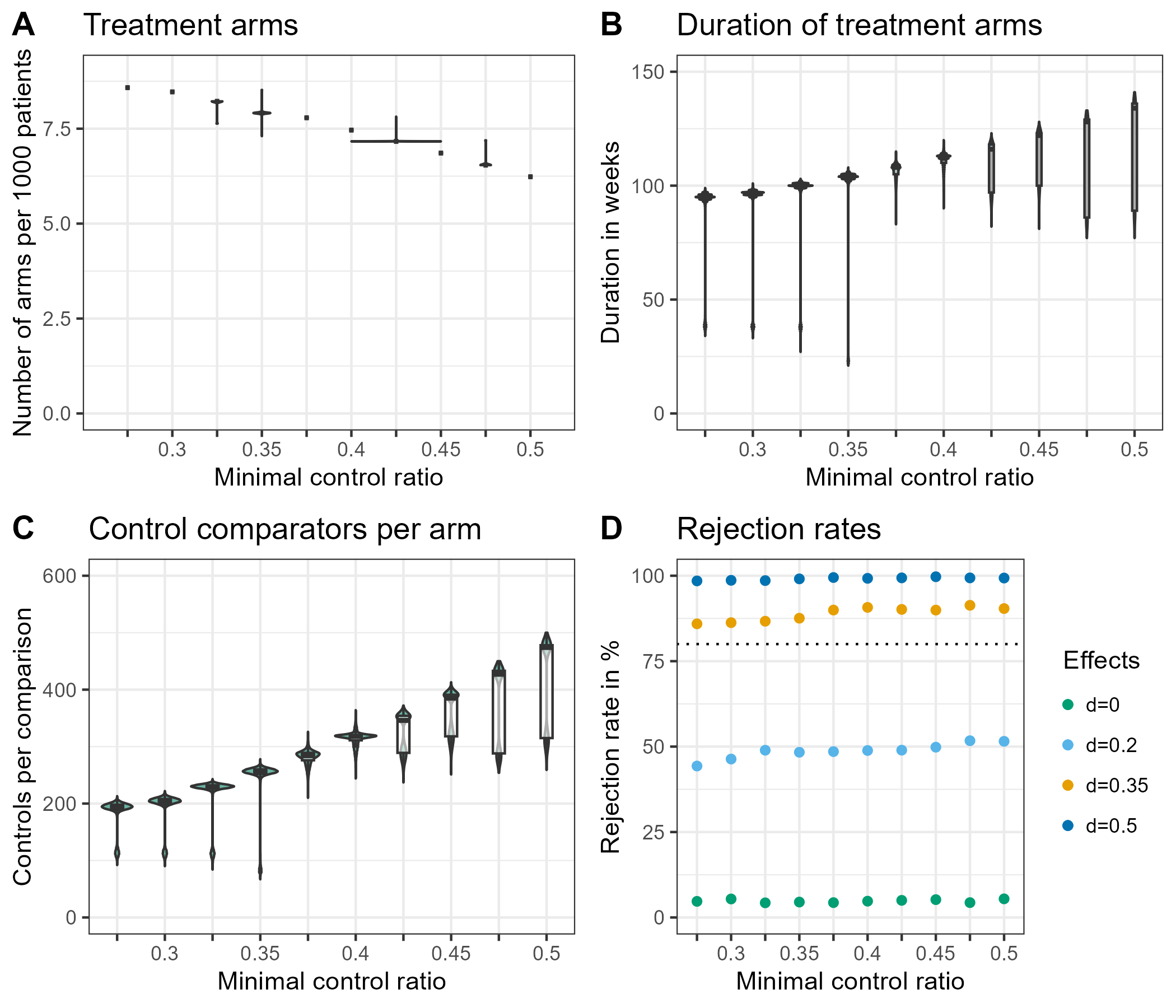}
    \caption{\csentence{Patients in the platform trial, on control, and on any treatment for different minimal control ratios.} As the platform trial allows for a maximum of 6 concurrent treatment arms, the left-most minimal control ratio of 0.275 corresponds to having no cap at all. The values are depicted as violin plots with integrated box plots.}
    \label{fig:cap_size}
\end{figure}

In order to assure that the allocation to control is always above a certain percentage one can set a minimal allocation to control. Due to the nature of Major Depressive Disorder we considered doing so as in this disease type a lower likelihood to receive placebo is linked to change of patients expectations and/or change in the patient population leading to an increase in the placebo response. From a clinical perspective having a control ratio of at least one-third is recommended for limiting the placebo response. The statistical implications of different minimal allocation caps to placebo were investigated by simulations and are summarized here. The per-arm sample size was set to 80, no interim analyses were conducted and an equal effect size distribution was assumed, i.e. the probability of each effect size set to$\theta_0 = \theta_{0.2} = \theta_{0.35} = \theta_{0.5} = 0.25$.
Figure \ref{fig:cap_size} shows the implications of different caps on the overall sample sizes, duration of treatment arms, size of control comparators, and rejection rates.
\\
Note that a minimal control ratio of 0.275 corresponds to having no cap at all but the standard $\sqrt{k}$ allocation because the platform trial only allows for a maximum of 6 concurrent treatment arms. A minimal control ratio of 0.5 on the other hand corresponds to the $k$ allocation, i.e., always half of all patients receiving the control (placebo) treatment.
A higher minimal control cap implies a decreasing number of patients on treatment and an increasing number on control while the overall sample size of the platform is more or less the same. The duration of treatment arms also increases with an increasing minimal control cap. Figure \ref{fig:cap_size}A and D show that it also implies increasing power while the number of arms that can be tested decreases.
\\
In our application, the platform trial should be designed for a phase II trial. In this setting, it is important to screen as many treatments as possible in a time as short as possible. For this reason, we selected 35\% as the minimal control ratio. This value is just above the one-third the clinicians recommended as a minimal boundary and allows for the most arms to be tested. Additionally, the power is still quite high and does not differ much from the power reached by implementing slightly higher minimal control ratios.

\clearpage

\section{Additional scenarios for the design options presented in the main paper}
In order to see the maximum benefit that can be reached by implementing a platform trial we present in the main paper results for platform trials running at maximum capacity, i.e. always 6 treatment arms running in parallel. For the comparison of platform trials to a series of two-arm trials a more realistic scenario for the availability of new treatment arms is presented. Here we show the results for this scenario in the other investigations.
Also, with exception to the futility section, in the main paper only results for the equal distribution of effect sizes are shown, i.e. $\theta_0 = \theta_{0.2} = \theta_{0.35} = \theta_{0.5} = 0.25$. However, there are many possible effect size distributions that can be assumed. Here we present results for the more pessimistic effect size scenario that is also applied in the futility section of the main paper, i.e. $\theta_0 = 0.5$,  $\theta_{0.2} = 0.3$, $\theta_{0.35} = 0.1$, $\theta_{0.5} = 0.1$.

\subsection{Additional scenarios for the selection of the allocation method}
\begin{figure}[h!]
  \centering
  \includegraphics[width=16cm]{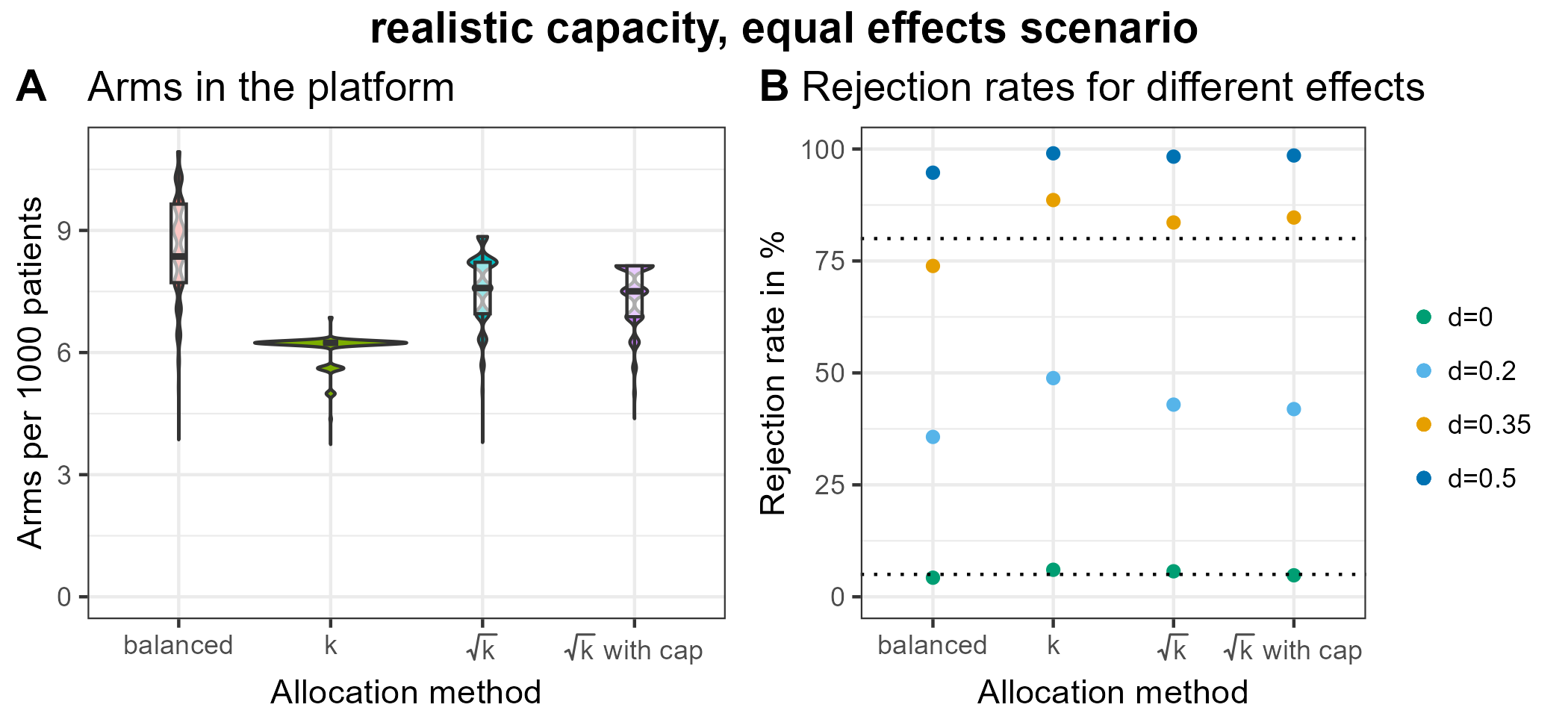}
    \caption{Arms per 1000 patients and rejection rates for different allocation methods in the scenario with more realistic availability of new treatment arms and equal effect size distribution. The targeted sample size per treatment arm was fixed at $N=80$. The minimal control cap for the $\sqrt{k}$ allocation was set to 35\%. A) depicts the number of arms that can be evaluated per 1000 patients in a corresponding platform trial. In B the percentage of rejected null hypotheses is depicted. It equals the type I error rate for $d = 0$ and the power for the other values of $d$. The type I error rate is always controlled at 5\%. This value is indicated by the lower dotted line. The higher dotted line highlights the 80\% mark.}
    \label{fig:alloc_s3}
\end{figure}

\begin{figure}[h!]
  \centering
  \includegraphics[width=16cm]{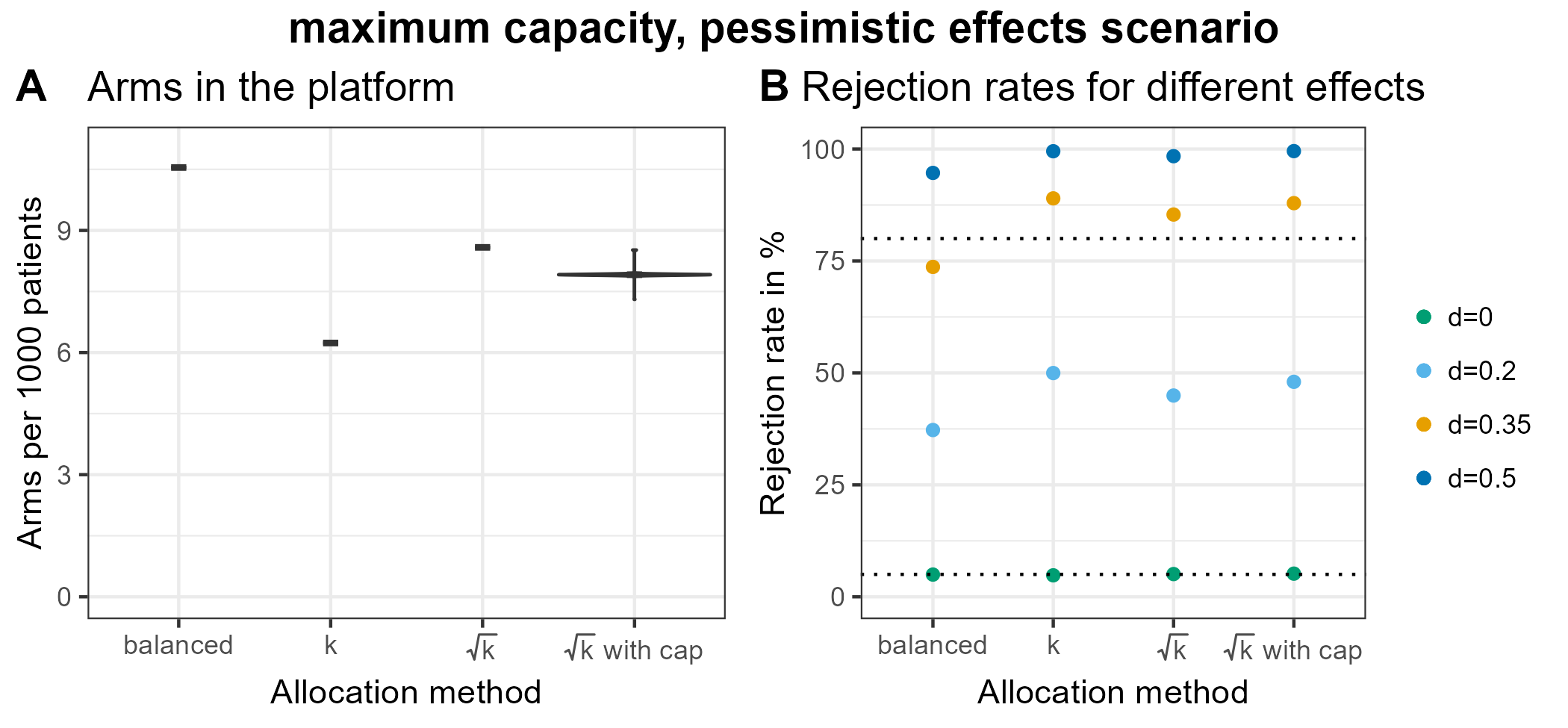}
    \caption{Arms per 1000 patients and rejection rates for different allocation methods in the scenario with the platform running at maximum capacity and a more pessimistic effect size distribution. The targeted sample size per treatment arm was fixed at $N=80$. The minimal control cap for the $\sqrt{k}$ allocation was set to 35\%. A) depicts the number of arms that can be evaluated per 1000 patients in a corresponding platform trial. In B the percentage of rejected null hypotheses is depicted. It equals the type I error rate for $d = 0$ and the power for the other values of $d$. The type I error rate is always controlled at 5\%. This value is indicated by the lower dotted line. The higher dotted line highlights the 80\% mark.}
    \label{fig:alloc_pes}
\end{figure}

\begin{figure}[h!]
  \centering
  \includegraphics[width=16cm]{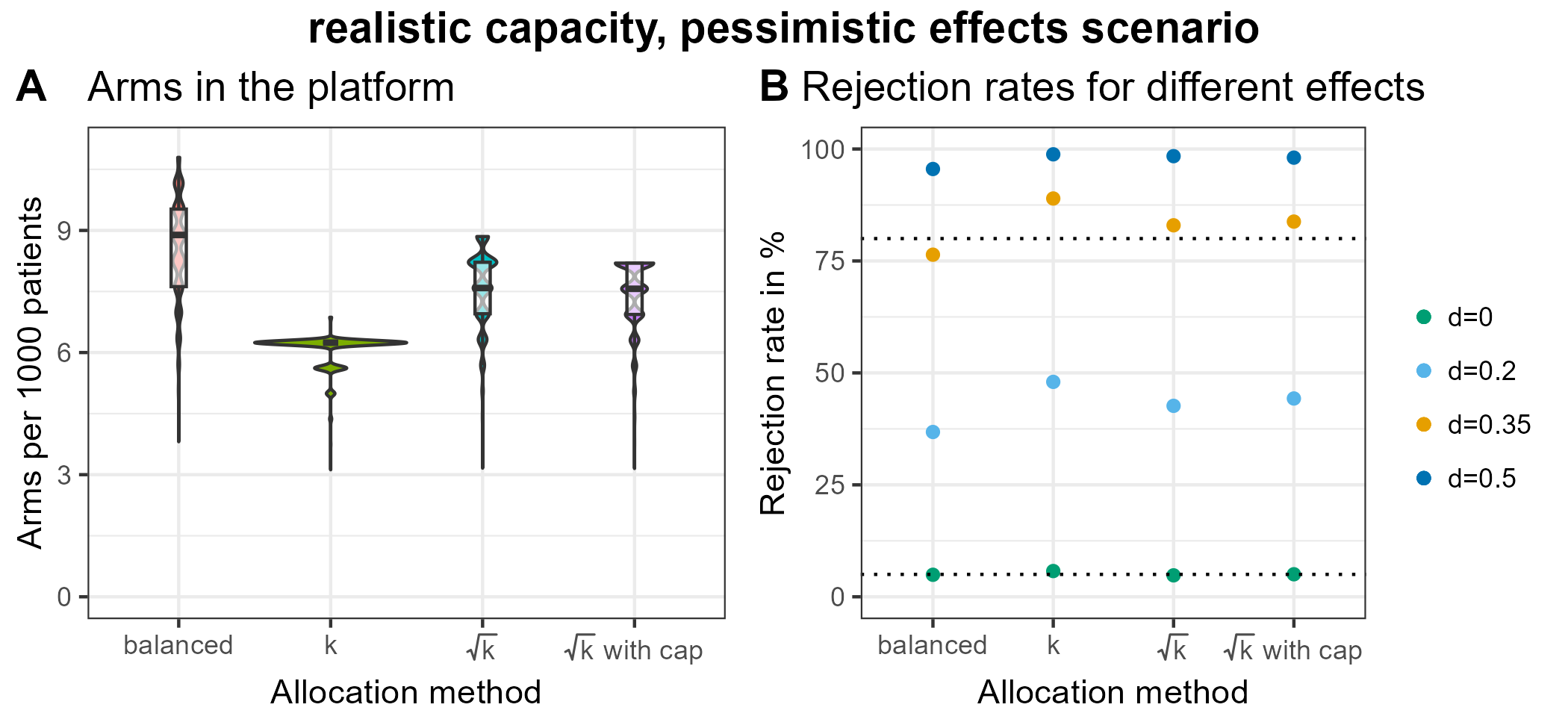}
    \caption{Arms per 1000 patients and rejection rates for different allocation methods in the scenario with more realistic availability of new treatment arms and a more pessimistic effect size distribution. The targeted sample size per treatment arm was fixed at $N=80$. The minimal control cap for the $\sqrt{k}$ allocation was set to 35\%. A) depicts the number of arms that can be evaluated per 1000 patients in a corresponding platform trial. In B the percentage of rejected null hypotheses is depicted. It equals the type I error rate for $d = 0$ and the power for the other values of $d$. The type I error rate is always controlled at 5\%. This value is indicated by the lower dotted line. The higher dotted line highlights the 80\% mark.}
    \label{fig:alloc_pes_s3}
\end{figure}

For the selection of allocation methods, the scenario of the platform running at maximum capacity and an equal effect size distribution is presented in the main paper. Here we show the results for the scenario with more realistic availability of new treatment arms and equal effect size distribution (Figure \ref{fig:alloc_s3}), the scenario with the platform running at maximum capacity and a more pessimistic effect size distribution (Figure \ref{fig:alloc_pes}), and the scenario with more realistic availability of new treatment arms and a more pessimistic effect size distribution (Figure \ref{fig:alloc_pes_s3}).
\clearpage

\clearpage

\clearpage

\subsection{Additional scenarios for the selection of the futility stopping rule}
\begin{figure}[h!]
  \centering
  \includegraphics[width=16cm]{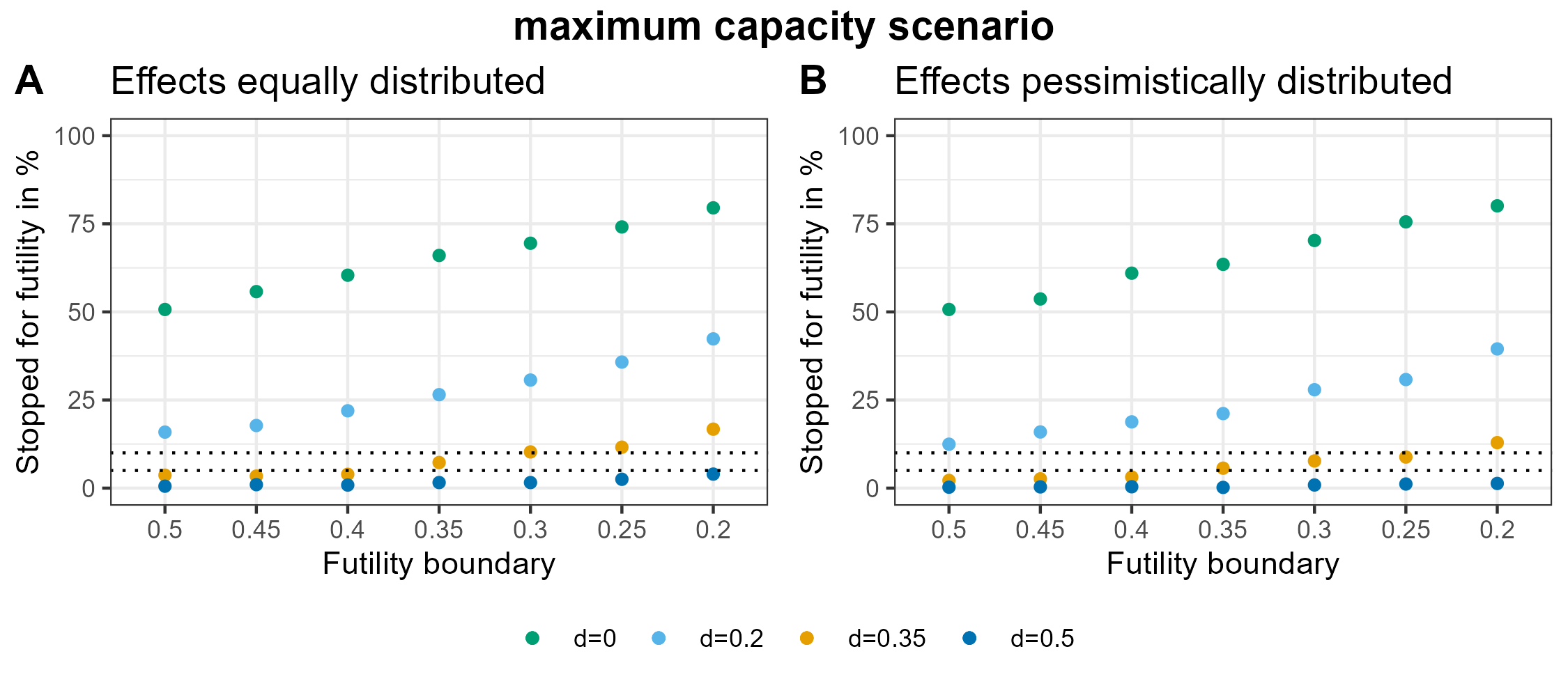}
    \caption{Rates of early stopping for futility in the scenarios with the platform running at maximum capacity. The lower dotted line indicates a level of 5\% and the higher dotted line highlights the 10\% mark. In A all effect sizes are assumed to be equally likely and in B we present results for a more pessimistic effect size scenario. }
    \label{fig:stopped}
\end{figure}

\begin{figure}[h!]
  \centering
  \includegraphics[width=16cm]{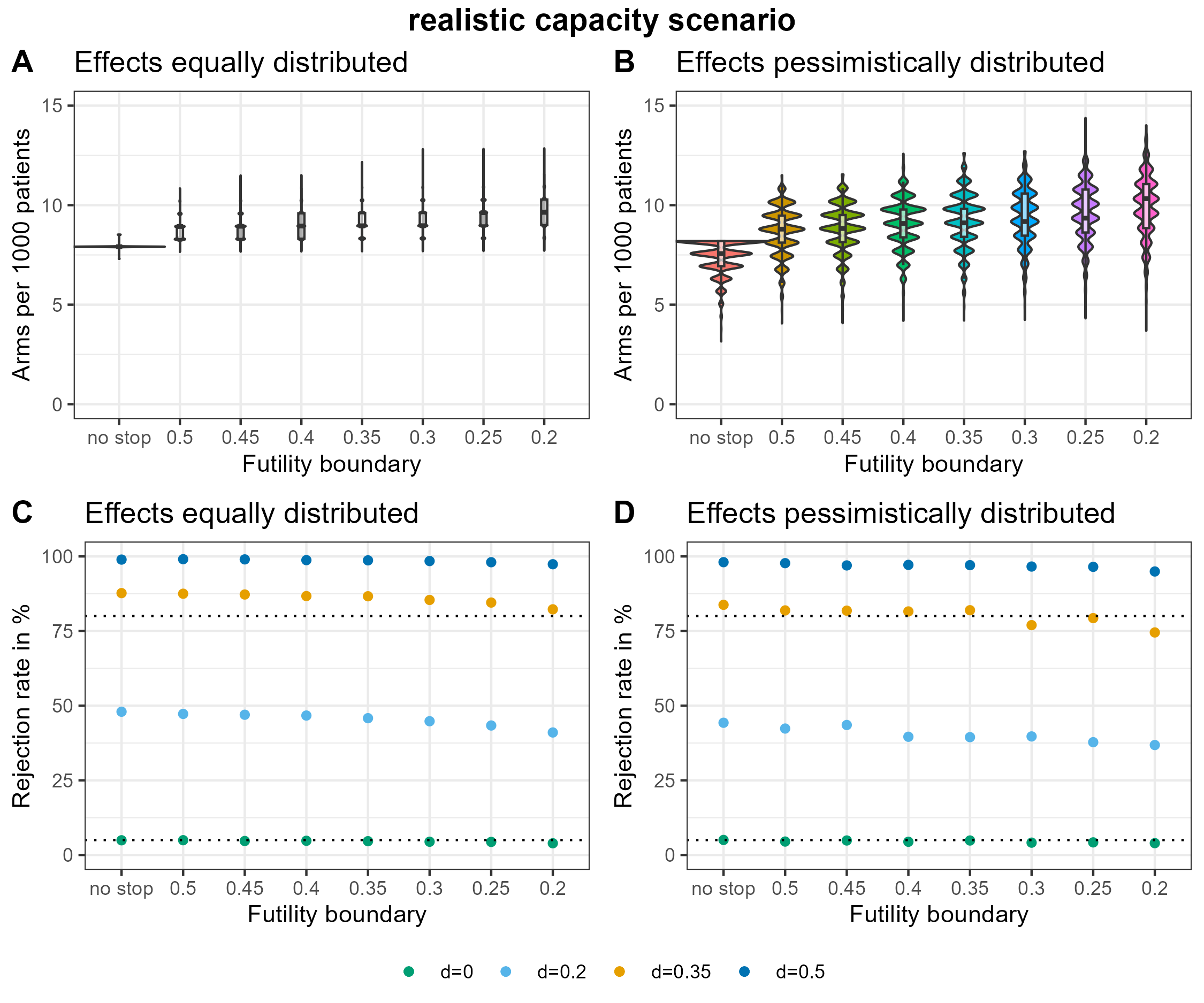}
    \caption{Rejection rates and standardized number of arms when implementing different futility rules in the scenarios with more realistic availability of new treatment arms. On the left side (A and C) all effect sizes are assumed to be equally likely. On the right side (B and D) we present results for a more pessimistic scenario. A and B give the standardized number of arms per 1000 patients in the platform trial. All arms are included in this number regardless which of the four different investigated effect sizes was allocated. In B and D the percentage of rejected null hypotheses is depicted. It equals the type I error rate for $d = 0$ and the power for the other values of $d$. The type I error rate is always controlled at 5\%. This value is indicated by the lower dotted line. The higher dotted line highlights the 80\% mark.}
    \label{fig:fut_s3}
\end{figure}

\begin{figure}[h!]
  \centering
  \includegraphics[width=16cm]{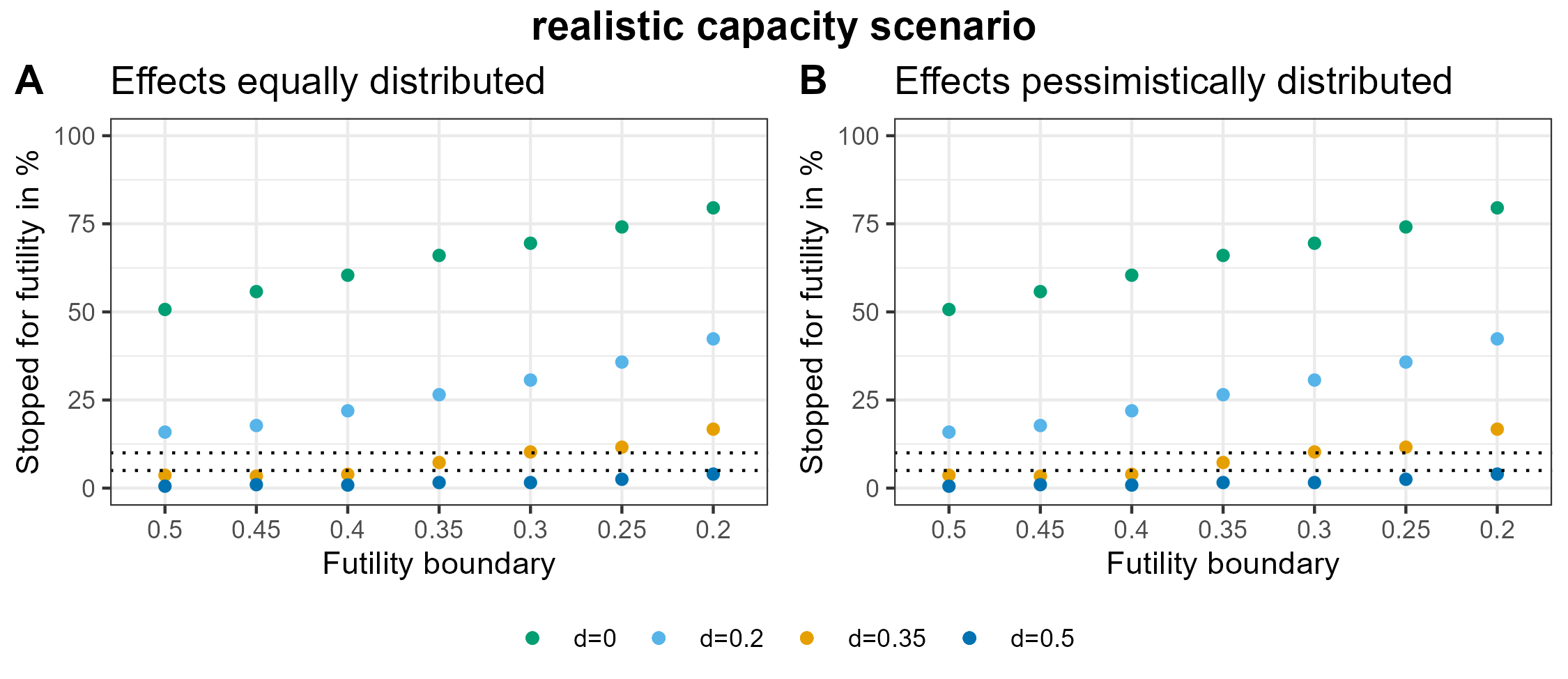}
    \caption{Rates of early stopping for futility in the scenarios with more realistic availability of new treatment arms. The lower dotted line indicates a level of 5\% and the higher dotted line highlights the 10\% mark. In A all effect sizes are assumed to be equally likely and in B we present results for a more pessimistic effect size scenario. }
    \label{fig:stopped_s3}
\end{figure}

For the selection of the futility stopping rule, both scenarios of the platform running at maximum capacity are presented in the main paper (i.e. with an equal effect size distribution and with a more pessimistic effect size distribution). Here we additionally show the rate of treatment arms that is stopped at the interim analysis in Figure \ref{fig:stopped}.
We also give the results for both scenarios with more realistic availability of new treatment arms in Figure \ref{fig:fut_s3} and \ref{fig:stopped_s3}).
\vspace{1cm}

\clearpage

\subsection{Additional scenarios for the selection of the per treatment arm sample size}
\begin{figure}[h!]
  \centering
  \includegraphics[width=16cm]{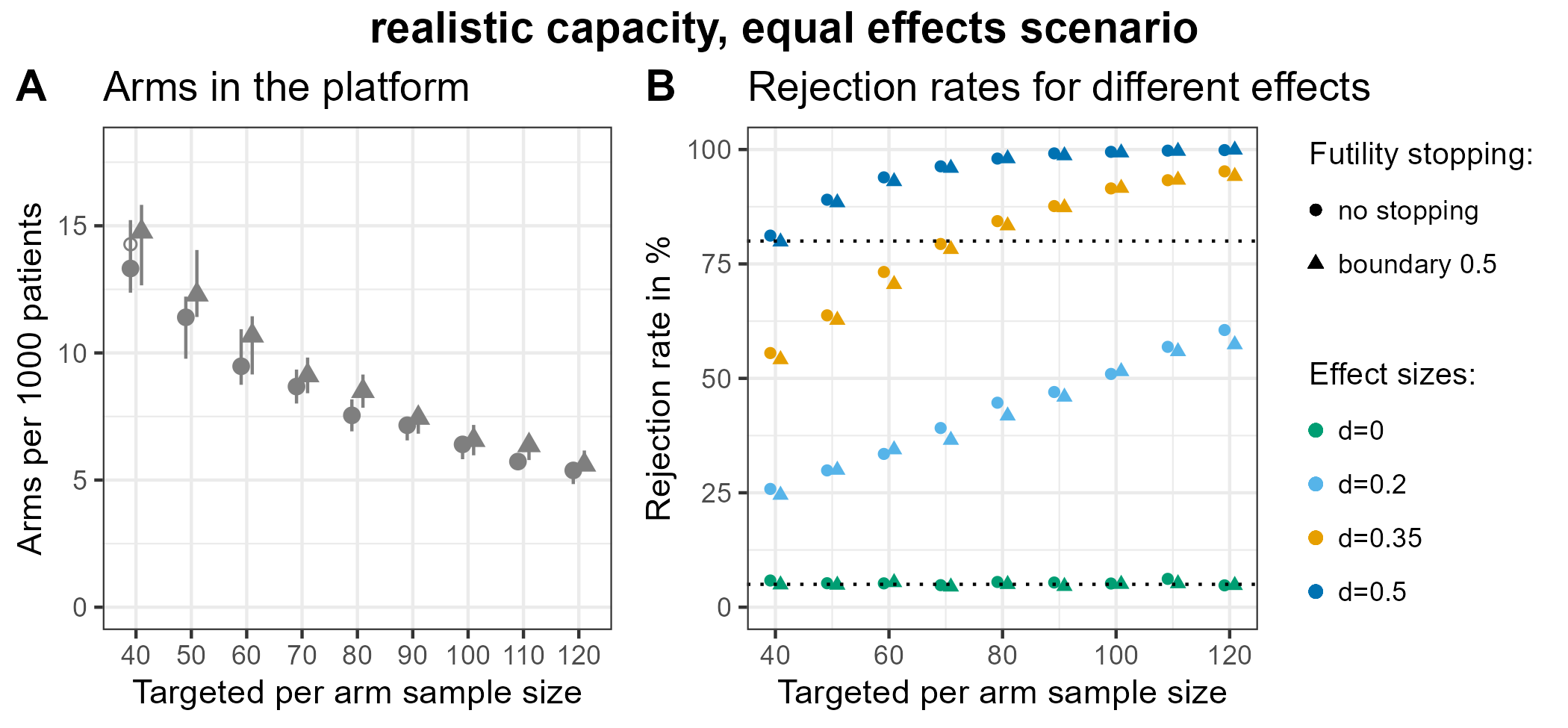}
    \caption{Rejection rates and standardized number of arms for different per-arm sample sizes for the scenario with more realistic availability of new treatment arms and equal effect size distribution. The circles give the values without the implementation of an interim analysis and the triangles the corresponding values when a futility boundary of 0.5 is applied. The sample size depicted on the x-axis was examined in steps of 10. The small variation in x direction is based on jittering for better readability. Figure A) shows the median number of arms per 1000 patients and the interquartile range. In B) the rejection rates are depicted stratified by the four different investigated effect sizes. The rejection rate equals the type I error rate for $d=0$ and the power for the other values of $d$. The type I error rate is always controlled at 5\%. This value is indicated by the lower dotted line. The higher dotted line highlights the 80\% mark.}
    \label{fig:n_s3}
\end{figure}

\begin{figure}[h!]
  \centering
  \includegraphics[width=16cm]{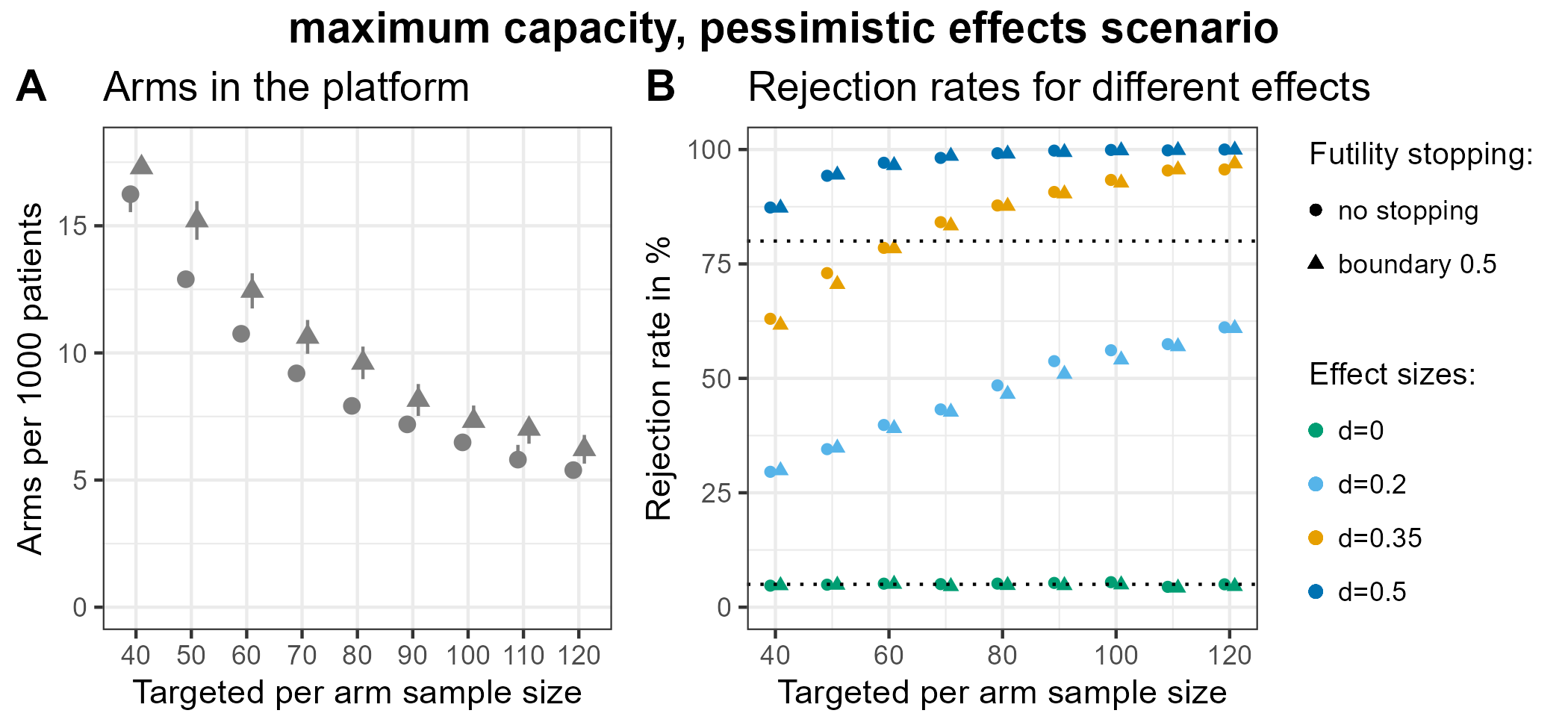}
    \caption{Rejection rates and standardized number of arms for different per-arm sample sizes for the scenario with the platform running at maximum capacity and a more pessimistic effect size distribution. The circles give the values without the implementation of an interim analysis and the triangles the corresponding values when a futility boundary of 0.5 is applied. The sample size depicted on the x-axis was examined in steps of 10. The small variation in x direction is based on jittering for better readability. Figure A) shows the median number of arms per 1000 patients and the interquartile range. In B) the rejection rates are depicted stratified by the four different investigated effect sizes. The rejection rate equals the type I error rate for $d=0$ and the power for the other values of $d$. The type I error rate is always controlled at 5\%. This value is indicated by the lower dotted line. The higher dotted line highlights the 80\% mark.}
    \label{fig:n_pes}
\end{figure}

\begin{figure}[h!]
  \centering
  \includegraphics[width=16cm]{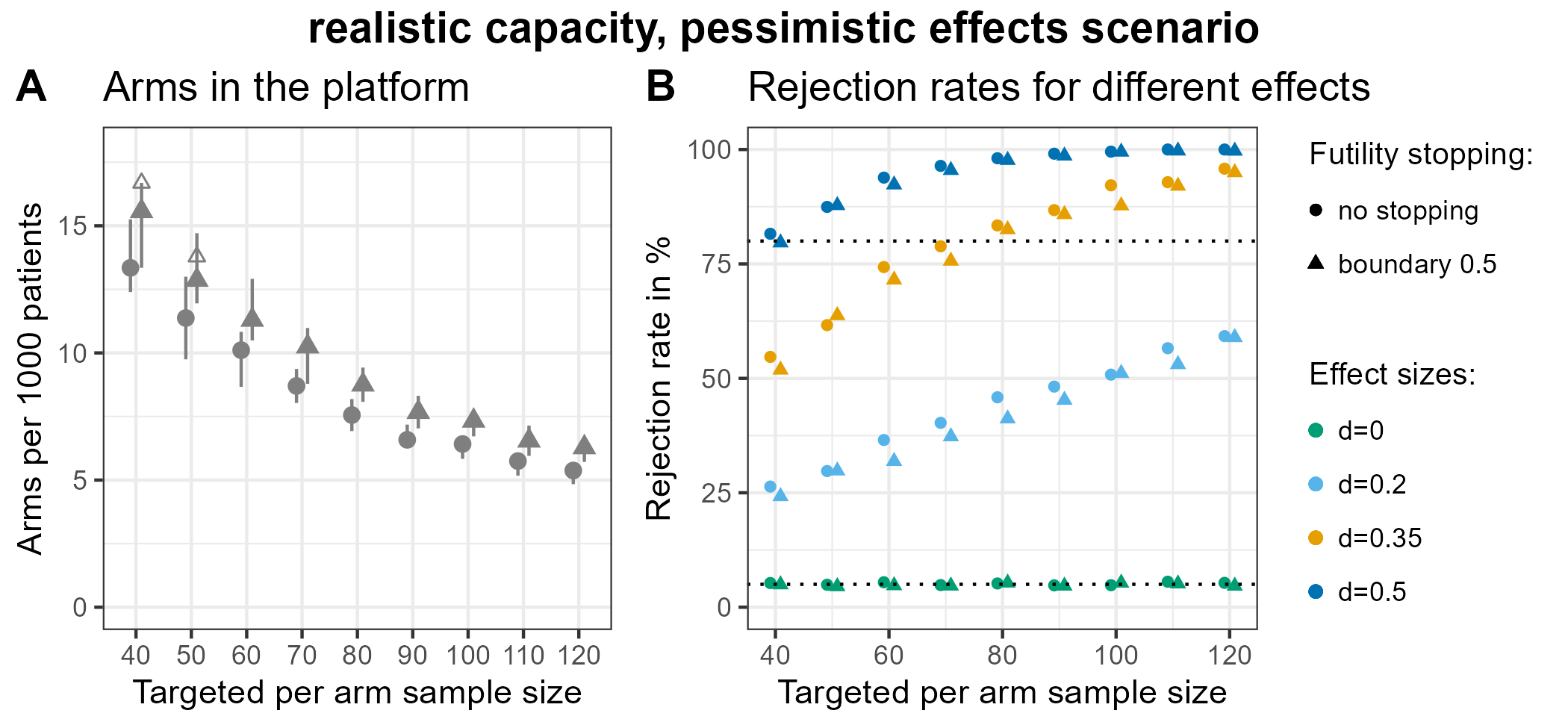}
    \caption{Rejection rates and standardized number of arms for different per-arm sample sizes for the scenario with more realistic availability of new treatment arms and a more pessimistic effect size distribution. The circles give the values without the implementation of an interim analysis and the triangles the corresponding values when a futility boundary of 0.5 is applied. The sample size depicted on the x-axis was examined in steps of 10. The small variation in x direction is based on jittering for better readability. Figure A) shows the median number of arms per 1000 patients and the interquartile range. In B) the rejection rates are depicted stratified by the four different investigated effect sizes. The rejection rate equals the type I error rate for $d=0$ and the power for the other values of $d$. The type I error rate is always controlled at 5\%. This value is indicated by the lower dotted line. The higher dotted line highlights the 80\% mark.}
    \label{fig:n_pes_s3}
\end{figure}

For the selection of the sample size per treatment arm, the scenario of the platform running at maximum capacity and an equal effect size distribution is presented in the main paper both without an interim analysis and with a futility boundary of 0.5. Here we show the results for the scenario with more realistic availability of new treatment arms and equal effect size distribution (Figure \ref{fig:n_s3}), the scenario with the platform running at maximum capacity and a more pessimistic effect size distribution (Figure \ref{fig:n_pes}), and the scenario with more realistic availability of new treatment arms and a more pessimistic effect size distribution (Figure \ref{fig:n_pes_s3}).
\vspace{1cm}
\clearpage
\clearpage

\clearpage
\vspace{1cm}

\subsection{Additional scenarios for the comparison to 2-arm trials}
\begin{figure}[h!]
  \centering
  \includegraphics[width=16cm]{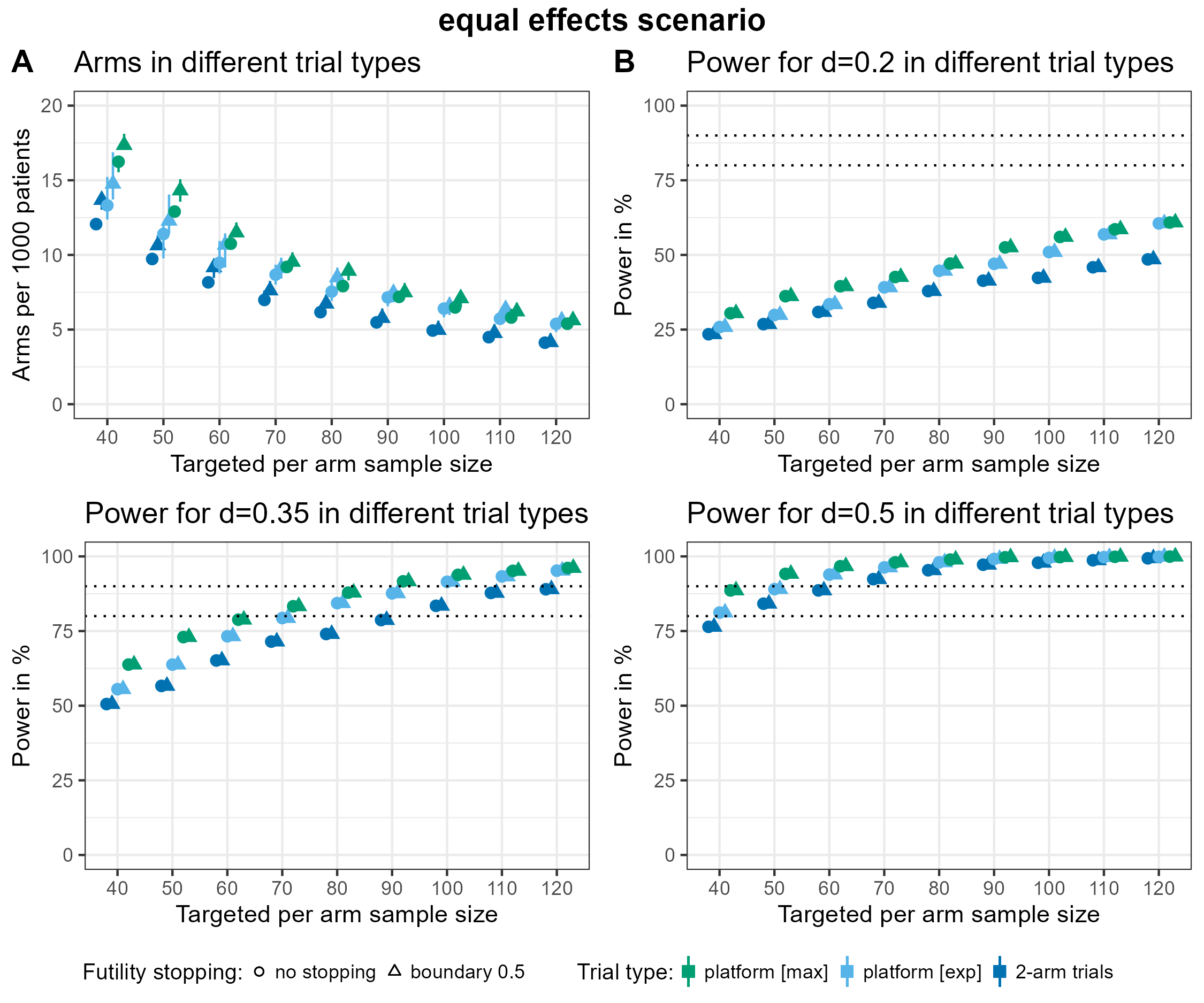}
    \caption{Comparison of operating characteristics in different trial types for the scenario with an equal effect size distribution. The circles give the values without the implementation of an interim analysis and the triangles the corresponding values when a futility boundary of 0.5 is applied. The sample size depicted on the x-axis was examined in steps of 10. The small variation in x direction is based on jittering for better readability. Figure A) shows the median number of arms per 1000 patients and the interquartile range for the three different trial types platform trial with maximum capacity utilization, platform trial with expected load in the MDD case, and the traditional approach with a series of individual 2-arm randomised controlled trials. B) gives the power for the same type of trials.}
    \label{fig:2arm}
\end{figure}

\begin{figure}[h!]
  \centering
  \includegraphics[width=16cm]{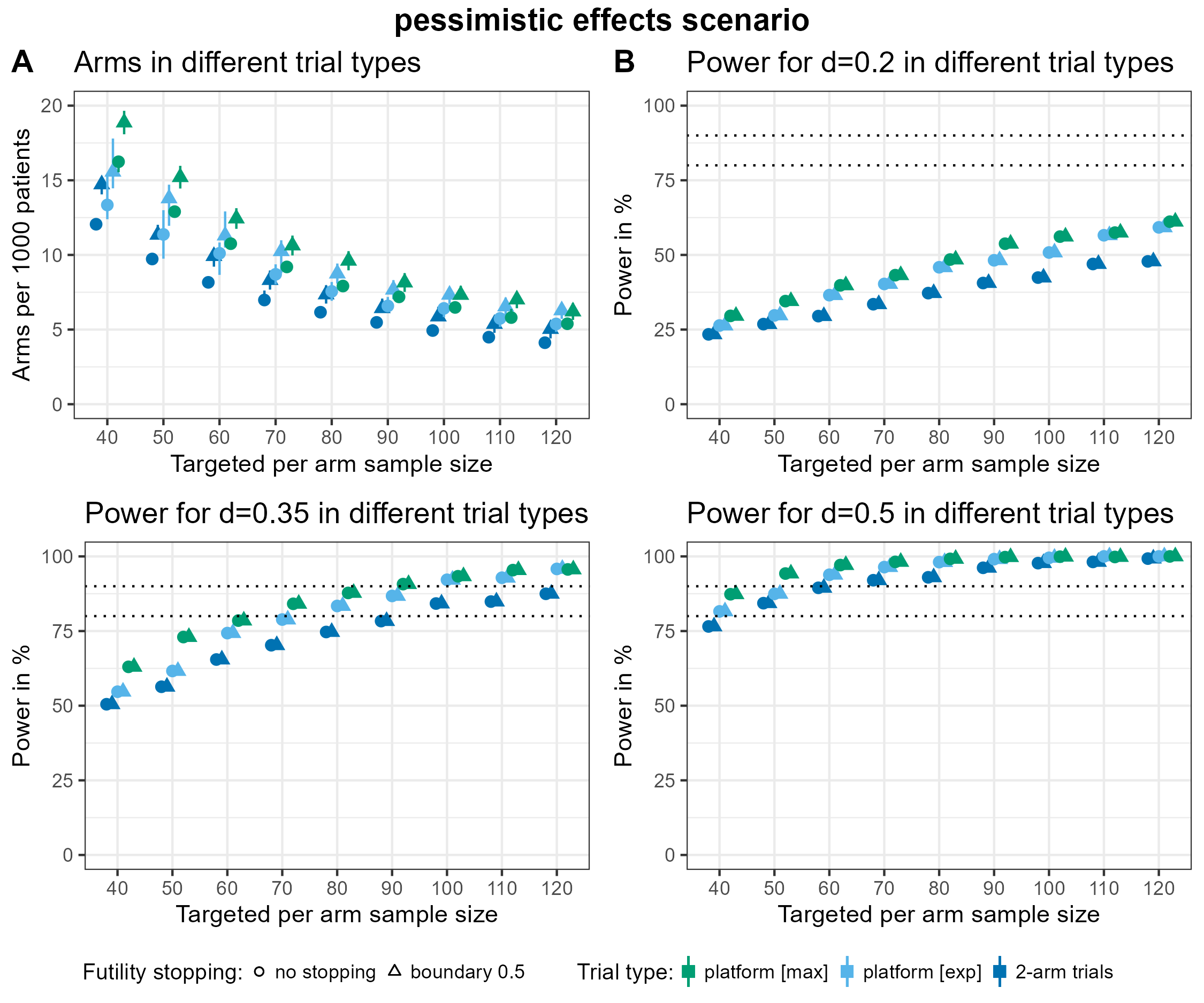}
    \caption{Comparison of operating characteristics in different trial types for the scenario with a more pessimistic effect sizes distribution. The circles give the values without the implementation of an interim analysis and the triangles the corresponding values when a futility boundary of 0.5 is applied. The sample size depicted on the x-axis was examined in steps of 10. The small variation in x direction is based on jittering for better readability. Figure A) shows the median number of arms per 1000 patients and the interquartile range for the three different trial types platform trial with maximum capacity utilization, platform trial with expected load in the MDD case, and the traditional approach with a series of individual 2-arm randomised controlled trials. B) gives the power for the same type of trials.}
    \label{fig:2arm_s3}
\end{figure}

In the section about the comparison to traditional 2-arm trials, a range of results for the scenario with an equal effect size distribution are presented in the main paper. Values for the platform running at maximum capacity and a more realistic availability of new treatment arms are depicted there. But the power is not explicitly shown for all different effect sizes and not all possible sample sizes per treatment arm are included. 
Here we present the full range of values for this scenario in Figure \ref{fig:2arm}. We also show these results for the scenario with a more pessimistic effect size distribution (Figure \ref{fig:2arm_s3}).

\clearpage

\section*{Acknowledgements}

The authors are grateful to the EU-PEARL investigators who contributed to the development of the MDD master protocol. 
The EU-PEARL MDD investigators are: 
Jelena Brasanac, Woo Ri Chae, Michaela Maria Freitag, Stefan Gold, Eugenia Kulakova, Christian Otte, Dario Zocholl, 
Francesco Benedetti, 
Witte Hoogendijk, 
Marta Bofill-Roig, Franz König, Martin Posch, 
Yanina Flossbach, 
Tasneem Arsiwala, Alexandra Bobirca,
Fernanda Baroso de Sousa, Pol Ibanez-Jimenez, Gabriela Perez-Fuentes, Toni Ramos-Quiroga,
Melissa Kose, Giulia Lombardi, Carmine Pariante, Luca Sforzini, Courtney Worrell 

